\documentclass[preprint,12pt]{elsarticle}
\usepackage{graphicx}
\usepackage{amssymb}
\usepackage{lineno}
\usepackage[usenames, dvipsnames]{xcolor}
\usepackage{graphicx} 
\usepackage{float} 
\usepackage{hyperref}
\usepackage{anysize}
\usepackage{amsmath,amsfonts,amsthm,bm}
\usepackage{ulem}

\usepackage{subfig}

\usepackage{natbib}

\setcitestyle{authoryear,open={(},close={)}}

\usepackage[colorinlistoftodos]{todonotes}

\begin{document}

\begin{frontmatter}

\title{ Approximate Bayesian inference for a ``steps and turns'' continuous-time random walk observed at regular time intervals}
\author{Sofia Ruiz-Suarez$^{1,2}$, Vianey Leos-Barajas$^3$, Ignacio Alvarez-Castro$^4$, Juan M.\ Morales$^1$}

\address{ \vspace{4mm} \small{$^1$INIBIOMA (CONICET-Universidad Nacional del Comahue), Quintral 1250, Bariloche, Rio Negro, Argentina} \\
 \small{$^2$Universidad de Rosario, Facultad de Ciencias Econ\'{o}micas, Bv. Oro\~{n}o 1261, Rosario,Argentina} \\
\small{$^3$ Department of Statistics, Iowa State University, Snedecor Hall, Ames, IA 50011, USA} \\ \small{$^4$ Universidad de la Rep\'{u}blica, Eduardo Acevedo 1139 Montevideo, UY 11200}}


\begin{abstract}

 The study of animal movement is challenging because it is a process modulated by many factors acting at different spatial and temporal scales. Several models have been proposed which differ primarily in the temporal conceptualization, namely continuous and discrete time formulations. Naturally, animal movement occurs in continuous time but we tend to observe it at fixed time intervals. 

To account for the temporal mismatch between observations and movement decisions, we used a state-space model where movement decisions (steps and turns) are made in continuous time. The movement process is then observed at regular time intervals.

As the likelihood function of this state-space model turned out to be complex to calculate yet simulating data is straightforward, we conduct inference using a few variations of Approximate Bayesian Computation (ABC). We explore the applicability of these methods as a function of the discrepancy between the temporal scale of the observations and that of the movement process in a simulation study.  We demonstrate the application of this model to a real trajectory of a sheep that was reconstructed in high resolution using information from magnetometer and GPS devices. 


Our results suggest that accurate estimates can be obtained when the observations are less than 5 times the average time between changes in movement direction.

 
The state-space model used here allowed us to connect the scales of the observations and movement decisions in an intuitive and easy to interpret way. Our findings underscore the idea that the time scale at which animal movement decisions are made needs to be considered when designing data collection protocols, and that sometimes high-frequency data may not be necessary to have good estimates of certain movement processes. 

\end{abstract}


\end{frontmatter}

\section*{Introduction}

\label{S:1}

The way in which animals move is of fundamental importance in ecology and evolution. It plays important roles in the fitness of individuals and in gene exchange \citep{nathan_movement_2008}, in the structuring of populations and communities \citep{turchin_quantitative_1998,matthiopoulos_establishing_2015,morales_building_2010}, and in the spread of diseases \citep{fevre_animal_2006}. 
The study of movement is challenging because it is a process modulated by many factors acting at different spatial and temporal scales \citep{gurarie_characteristic_2011, mevin_b._hooten_animal_2017}. 
In general, the process of animal movement occurs in continuous time but we observe individual locations at time intervals dictated by logistic constrains such as battery life. It is necessary to be aware of this fact in order to avoid drawing conclusions about the movement process that depend on the time scale in which the observations were taken.

In this context, state-space models provide a convenient  tool for movement data analysis \citep{patterson_state-space_2008}. The main idea is to estimate the latent movement process given the observational process. Thus, they consist of two stochastic models: a latent model and an observation one. The first describes the state of the animal (it could be the location, behaviour, etc) and the second one describes the observation of the state, possibly with some measurement error. Several state-space models have been proposed to model animal movement differing primarily in the temporal conceptualization of the movement process, namely discrete and continuous time formulations \citep{mcclintock_when_2014}. On one hand, discrete-time models describe movement as a series of steps and turns (or movement directions) that are performed at regular occasions \citep{morales_extracting_2004,jonsen_robust_2005,mcclintock_general_2012}. Typically in these models the temporal scales of both the latent and the observation process are the same. Thus, the observation times coincide with the times in which the animal are assumed to make movement decisions. The advantage of this approach is that it allows the dynamics involved in the movement process to be conceptualized in a simple and intuitive way, which facilitates implementation and interpretation. On the other hand, continuous-time models have been proposed \citep{blackwell_random_1999,johnson_continuous-time_2008,harris_flexible_2013} in which the movement process is defined for any time and expressed through stochastic differential equations that account for the dependence between successive locations. The observation process is then truly independent from the movement process and does not need to be recorded at regular time intervals. The continuous-time approach has the advantage of being more realistic and that the inference is not affected by the choice of scale of observation. The main drawback is probably in the interpretation of instantaneous movement parameters (e.g., those related to Ornstein-Uhlenbeck processes and other diffusion models). In such manner both approaches have advantages and disadvantages, the discrete-time models are more intuitive and easy to interpret but can be considered less realistic than continuous ones \citep{mcclintock_when_2014}. 


In this work, we present a state-space model that formulates the movement process in continuous time and the observation process in discrete time (regular intervals). As a compromise between the ease of interpretation of models based on steps and turns and the realism of continuous-time models, we use a random walk where the movement decisions (steps and turns) are made in continuous time. The movement process is then observed at regular time intervals. In this model there are two different time scales: one for the latent process and one for the observation process. The advantage here is that this model allows us to differentiate between the times in which the animals make movement decisions and the times in which the observations are made. 
One challenge faced with the proposed model formulation is that the resulting likelihood function seems to be computationally intractable.
However, simulating the movement and observation process is straightforward suggesting that likelihood-free methods such as Approximate Bayesian Computation (ABC) could be useful \citep{beaumont_approximate_2010, csillery_approximate_2010}. 


Here we describe, formalize, and expose the possible complications of a state-space movement model with two different temporal scales. We use stochastic simulations to evaluate the ability of three ABC techniques to recover the parameter values driving the movement process. Keeping in mind the ecological purpose behind implementing such a model, we assess the quality of these estimations with regard to the relationship between the two temporal scales. Finally, we apply the model to a high resolution trajectory of sheep to evaluate the performance of the ABC inference with real data.

\section*{Methods}
\subsection*{Movement Model with Random Time between movement decisions}
The general structure of the model is based on a correlated random walk. An individual moves in a certain direction for a certain period of time, and then it makes a turn and starts moving in a new direction for another period of time. Since in practice the path of an animal is usually observed at particular sampling occasions, we consider that the observation process occurs at regular time intervals. Therefore the observation process lies in the location of the individual every time $\Delta t$. As a simplification, we assume that there is no observation error. Our movement model is a form of the velocity jump process \citep{othmer1988models} where the speed of movement during the active phase is constant, and the temporal scale of the waiting time of the reorientation phase is considered instantaneous. Assuming constant movement speed, let the variable $M_i$ describe the position of the latent process at step $i$, presented in x-y coordinates, i.e. $M_i=(\mu_{i,1},\mu_{i,2})$ where $i$ represents an index of the time over the steps for $i=0,\dots,N_{steps}$. Given, $\mu_{0,1}=0$ and $\mu_{0,2}=0$, we have for $i=1,...,N_{steps}$ that,

\begin{equation} \label{eq:latent.model}
\begin{array}{cc}

\mu_{i,1}= & \mu_{i-1,1}+cos(\phi_{i-1})t_{i-1}   \\
 \mu_{i,2}= & \mu_{i-1,2}+sin(\phi_{i-1})t_{i-1} \\ 
\phi_i= & \sum_{k=1}^i \omega_i
\end{array}    
\end{equation}
where $t_i$  is the duration of step $i$, and $\omega_i$ is the turning angle between steps $i$ and $i+1$, so that $\phi_i$ represents the direction of the step $i$.  Each $t_i$ is assumed to be independently drawn from an exponential distribution with parameter $\lambda$ and each $\omega_i$ from a von Mises distribution with a fixed mean $\nu = 0$ and parameter $\kappa$ for the concentration. While the model can be extended to allow $\kappa$ and $\lambda$ to depend on the landscape, environment, or animal behaviour, for this work we only consider the initial case as an starting point. 

Next, we define the observation process and its links with the latent movement process. Let $O_j = (o_{j,1},o_{j,2})$ denote the position of observation $j$ in x-y coordinates, with $j=0,\dots,N_{obs}$. A second index $j$ is used for the time over the observations. Equations \eqref{eq:obs} show the relationship between observational and latent process. For this, it is necessary to determine the number of changes in direction that occurred before a given observation, we define $N_j$ as the number of steps (or changes in direction) that the animal took from time $1$ to time $j\Delta t$.
\begin{center}
$O_0=M_0$
\end{center}

\begin{equation}\label{eq:obs}
\begin{array}{cc}
o_{j,1}= &\mu_{N_{j},1}+cos(\phi_{N_j})\left(j\Delta t-\displaystyle\sum_{k<N_j-1} t_k  \right)=h_1(M_{0:N_j}) \\
o_{j,2}= &\mu_{N_{j},2}+sin(\phi_{N_{j}})\left(j\Delta t -\displaystyle\sum_{k<N_j-1} t_k  \right)=h_2(M_{0:N_j})
\end{array}
\end{equation}

\begin{figure}[ht]
 \centering
 \includegraphics[width=0.4\textwidth]{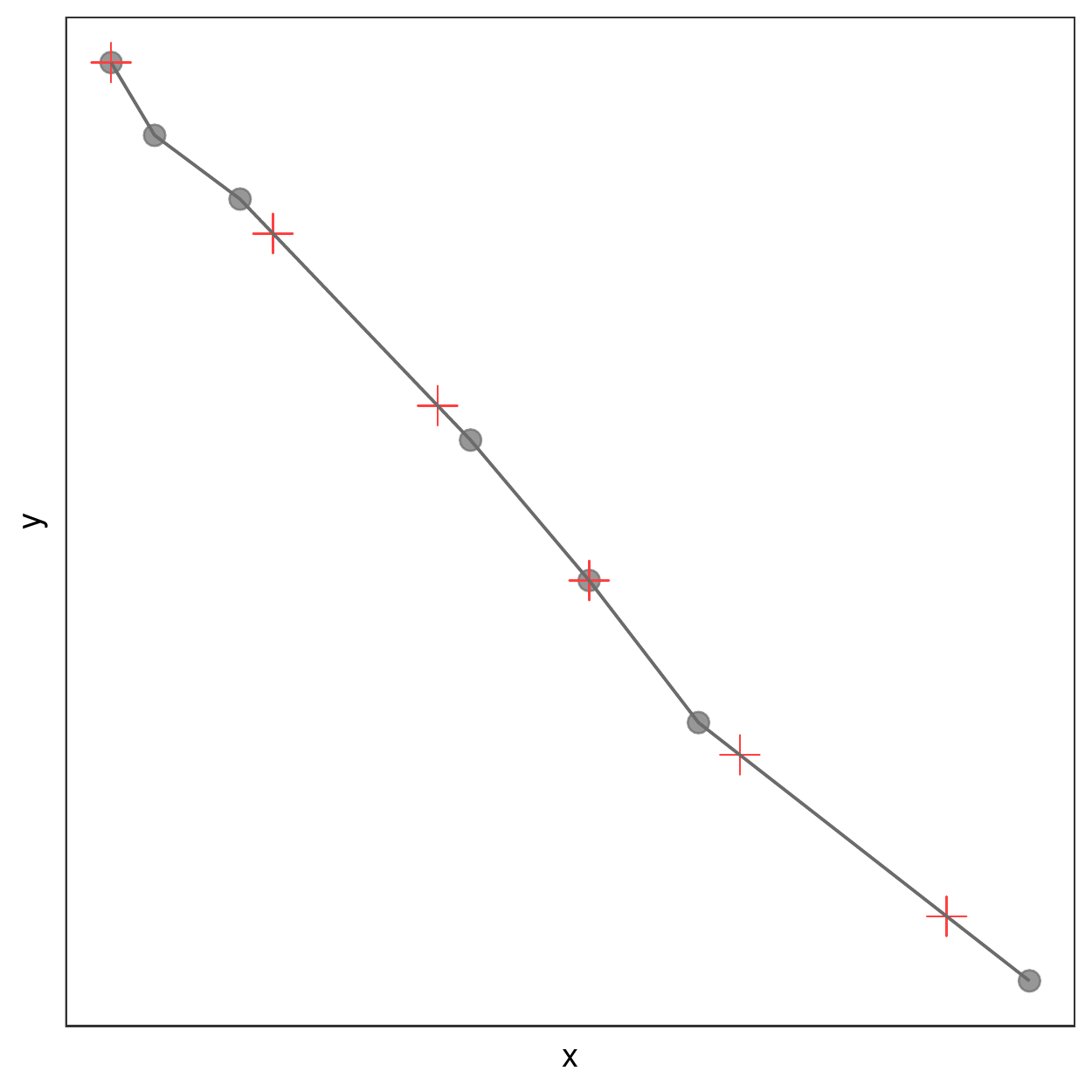}
 \caption{Example of few steps. With grey dots showing the direction change points and red crosses the observation points. Before the first and the second grey points (the initial one does not count, $i=0$) there is only one red cross, then $N_1=N_2=1$. Before the the third grey point there are three red crosses, then $N_3=3$, etc.}
 \label{ej_trayectoria}
\end{figure}

Then, $O_j=h(M_{0:N_j})$, a function of all positions $M_i$ from $i=0$ to $i=N_j$ with $M_{0:D}=(M_0,M_1,M_2,\dots,M_D)$. 
Note that $N_j$ is the maximum index such that the sum over all the duration times of the steps less or equal to it are at most $j\Delta t$, is possible to express it as $N_j=\max  \left\{ m / \sum_{s=0}^m t_s\leq j\Delta t \right\}$.  Therefore, the location $j$ is the last location of the latent process given by $N_j$ plus the difference between $j \Delta t$ and the time at which the step $N_j$ was produced in the direction $\phi_{N_j}$.
To better understand this relationship consider a minimal example of a few steps (Figure \ref{ej_trayectoria}). Lets assume the duration of steps and turning angles of Table \ref{tabexample} and $\Delta t=0.5$. In that case $N_1=1$, because $t_0=0.2\leq 1 \Delta t$, $t_0+t_1=0.4\leq 1 \Delta t$ but $t_0+t_1+t_2=1.1\nleq 1 \Delta t$. With the same reasoning $N_2=1$, $N_3=3$, $N_4=4$, $N_5=4$, etc.

\begin{table}
\centering
\caption{Values for the turning angles and duration of steps for the example of Figure \ref{ej_trayectoria}}
\begin{tabular}{|c|c|c|c|c|c|c|c|}
\hline
\textbf{Duration of steps ($t_i's$)} & 0.2& 0.2& 0.7& 0.4& 0.4& 0.8 \\ 
\hline
\textbf{Turning angle ($\omega_i's$)} &  0.32 & 5.65 & 5.81 & 0.02 & 0.11 & 5.81\\
\hline
\end{tabular}
\label{tabexample}
\end{table}

\subsection*{Expression for the likelihood function}

The likelihood function for the defined process is a function of the number of changes in direction that occurred, $\{N_j\}_{j=1}^{N_{obs}}$, and the parameters that determine the underlying movement process. 

To construct this expression recall that $O_j$ is a function of all positions $M_i$ from $i=0$ to $i=N_j$, $O_j=h(M_{0:N_j})$ (Eq. \ref{eq:obs}). We first suppose that we know the number of changes of direction that the animal took between consecutive observations, i.e. we know $N_j \forall j$. Therefore, we can express the likelihood as a function of the underlying movement parameters given both the observational process, $\{o_j\}_{j=1}^{N_{obs}}$, and the number of changes in direction, $\{N_j\}_{j=1}^{N_{obs}}$, known as the \textit{complete-data} likelihood. 

\begin{linenomath*}
\begin{equation*}
\begin{split}
L(\kappa,\lambda ; M, O ) & = P\left(O_0=o_0,O_1=o_1,\cdots,O_{N_{obs}}=o_{N_{obs}}\right)\\
& = P\left(h(M_{0:N_1})=o_1,h(M_{0:N_2})=o_2,...,
h(M_{0:N_{N_{obs}}})=o_{N_{obs}} \right)
\end{split}
\end{equation*}
\end{linenomath*}

In order to derive the complete-data likelihood, $L(\kappa, \lambda; M, O )$, it is necessary to obtain the distributions of $M_i$ (Eq. \ref{eq:latent.model}) and of $O_j = h(M_{0:N_j})$ (Eq. \ref{eq:obs}), which are not available in closed form. 
For derivation of the marginal likelihood, $L(\kappa, \lambda)$, it is further necessary to integrate over all the possible values of $N_j$, by determining $P(N_j=r)$ for $r \in  \mathbb{N}$, which can be, in principle, infinite.  
Obtaining the expression for and evaluation of the likelihood results to be a complex task. Likelihood-free methods that allow one to circumvent the need to evaluate the likelihood, such as ABC, have proven to be useful in these cases. 
es. 

\subsection*{Inference Using Approximate Bayesian Computation}

Approximate Bayesian Computation (ABC) is a family of simulation-based techniques to obtain posterior samples in models with an intractable likelihood function. In recent years, ABC has become popular in a diverse range of fields \citep*{Scott2018} such as molecular genetics \citep*{marjoram_modern_2006}, epidemiology \citep*{tanaka_using_2006,mckinley_inference_2009,lopes_abc:_2010}, evolutionary biology \citep{bertorelle_abc_2010,csillery_approximate_2010,baudet_cophylogeny_2015}, and ecology \citep{beaumont_approximate_2010, siren_assessing_2018}. This approach is also useful when the computational effort to calculate the likelihood is large compared to that of the simulation of the model of interest. The likelihood function described earlier turns out to be complex to calculate, yet it is easy to simulate trajectories from the statistical model, based on independent draws from exponential and von Mises distributions combined with the observations at regular time intervals. 

Let \bm{$\theta$} denote the vector of parameter of interest and \bm{$y$} denote the observed data. The posterior distribution $p(\bm{\theta}\mid \bm{y})$ is proportional to the product of the prior distribution $p(\bm{\theta})$ and the likelihood function $p(\bm{y}\mid \bm{\theta})$

\begin{center}
$p(\bm{\theta}\mid \bm{y})\propto p(\bm{\theta})p(\bm{y}\mid\bm{\theta})$ \end{center}

 The basic idea of ABC methods is to obtain simulations from the joint distribution, $p(\bm{y},\bm{\theta})$ and retain the parameter values that generate synthetic data close to the observed data, $\bm(y)$. In this way, ABC methods aim to replace the likelihood function with a measure of similarity between simulated data and actual data.

The rejection algorithm is the simplest and first proposed method to perform ABC \citep*{tavare_inferring_1997, pritchard_population_1999}. It can be described as follows: 
\begin{enumerate}
 \item 	Compute a vector of summary statistics with observed data, $S(\bm{y})$.

 \item Simulate parameters $\bm{\theta_{*}}$ sampled from $p(\bm{\theta})$ and synthetic data  $\bm{y_{*}}$ sampled from $p(.\mid\bm{\theta_{*}})$.
  \item Compute a vector of summary statistics with simulated data, $S(\bm{y}_*)$.
\item $\bm{\theta}_*$ is accepted as a posterior sample, if $\rho(S(\bm{y}_* ),S(\bm{y}))<\delta$, for some distance measure $\rho$ and threshold $\delta$. \item Repeat 2-4 $\bm{K}$ times. 
\end{enumerate}

The above rejection algorithm produces samples from $p(\bm{\theta}\mid\rho(S(\bm{y}),S(\bm{y}_* ))<\delta$ which is an approximation of $p(\bm{\theta}\mid \bm{y})$. In particular, when the summary statistics are sufficient or near-sufficient for $\rho$, the approximate posterior distribution converges to the true posterior distribution as $\delta$ goes to $0$ {\citep*{marjoram_markov_2003}}. Instead of selecting a value for $\delta$, it is a common practice to set a threshold $\epsilon$ as a tolerance level to define the proportion of accepted simulations.
For a complete review of ABC methods and techniques see \citep*{csillery_approximate_2010, beaumont_approximate_2010, handbook_abc}.

For this work we consider two regression-based correction methods. These implement an additional step to correct the imperfect match between the accepted and observed summary statistics. One of these use local linear regression 
\citep*{beaumont_approximate_2002}, and the other is based on neural networks \citep*{blum_non-linear_2010}. To make the correction, both methods use the regression equation given by 
\begin{center}
$\theta_i= m(S(y_i))+\xi_i$    
\end{center}

where $m$ is the regression function and the $\xi_i$’s are centered
random variables with equal variance. For the linear correction $m$ is assumed to be linear function and for the neural network correction $m$ is not necessary linear. 
A weight $K[d(S(y_i),S(y_0))]$ (for $K$ a statistical kernel) is assigned to each simulation, so those closer to the observed summary statistics are given greater weight. 
The $m$ and $\xi$ values can be estimated by fitting a linear regression in the first case and a feed-forward neural network regression in the second case. Then, a weighted sample from the posterior distribution is obtained by considering  $\theta_{i}^{corr}$ as follows

\begin{center}
$\theta_{i}^{corr}= \hat m(S(y_0))+\hat\xi_i$    
\end{center}

where $\hat m(.)$ is the estimated conditional mean and the $\hat\xi_i$s
are the empirical residuals of the regression.

After a preliminary analysis, in which $20$ summary statistics were assessed, we choose four that characterize the trajectories according to parameter values. Looking for summaries that capture diverse features of the movement, we plotted the proposed summaries against known parameters and decided to keep those summary statistics that changed monotonically with parameters values. The plots of all the summaries assessed are provided in the Appendix (Supporting Information). 
Finally, the four selected summaries were: a point estimator for $\lambda$, calculated as the inverse of the observed average step length (where an observed step is the distance between positions of consecutive observed times); a point
estimator for $\kappa$, calculated as the inverse function of the ratio of the first and zeroth order Bessel functions of the first kind evaluated at the mean of the cosine of the observed turning angles (where observed turning angles were determined as the difference between consecutive directions in the observations); the standard deviation of the observed turning angles and lastly, the standard deviation of the observed step lengths (Table \ref{table3}). 

\begin{table}
\centering
\caption{Summary statistics selected to fit the model using the three ABC
algorithms}
\begin{tabular}{|p{4.5cm}|p{8.5cm}|}
\hline
\textbf{Summary Statistic} & \textbf{Formula} \\[0.3ex] 
\hline
(1) Point estimate for $\lambda$ &
$1/\displaystyle\sum_{j=0}^{N_{obs}}t_{obs,j}$, 

with

$t_{obs,j}=\displaystyle\sqrt{(o_{j+1,1}-o_{j,1})^2+(o_{j+1,2}-o_{j,2})^2}$\\
\hline
(2) Point estimate for $\kappa$ & 

$\displaystyle A^{-1}\left(\frac{\displaystyle\sum_{j=1}^{N_{obs}} \cos(\omega_{obs,j})}{N_{obs}}\right)$

Where $\omega_{obs,j}=\arctan(\frac{o_{j+1,1}-o_{j,1}}{o_{j+1,2}-o_{j,2}})$

and $A(x)=\frac{I_1(x)}{I_0(x)}$
\citep*{kurt_hornik_maximum_2014}

\\
\hline
(3) Standard deviation of the turning angle & 
$\displaystyle\sqrt{\frac{\displaystyle\sum_{j=0}^{N_{obs}}\omega_{obs,j}-\bar{\omega_{obs}}}{N_{obs}-1}}$
\\
\hline
(4)Standard deviation of the step length &  
$\displaystyle\sqrt{\frac{\displaystyle\sum_{j=0}^{N_{obs}}t_{obs,j}-\bar{t_{obs}}}{N_{obs}-1}}$

\\
\hline
\end{tabular}
\label{table3}
\end{table}

We used the R package 
``abc"  (\citet*{csillery_abc:_2012},
http://cran.r-project.org/web/packages/abc/index.html) to perform 
the analysis. This package uses a standardized Euclidean distance to compare the observed and simulated summary statistics.
We present results for the two regression-based correction methods and for the basic rejection ABC method.

\subsection*{Simulations}
We did two simulation experiments. First we assessed the performance of the three ABC methods for our model. Then, we evaluated how well these methods approximate posterior probabilities depending on the relation between the temporal scales of simulated trajectories and their observations.   
For both experiments we used a set of one million simulated trajectories,  
 with parameters $\kappa$ (dispersion parameter for the turning angles) and $\lambda$ (parameter for the times between changes of direction) drawn from the priors  $p(\kappa)\sim U[0,100]$ and $p(\lambda)\sim U[0,50]$. The number of simulated steps was such that all the trajectories had at least $1500$ observations. All trajectories were observed at regular times of  $\Delta t=0.5$.

\subsubsection*{Assessment of the inference capacity of the ABC methods}\label{exp1}

We assessed the performance of the three ABC versions: simple rejection, rejection corrected via linear regression and rejection corrected via neural network. For different values of threshold ($\epsilon$) and for each algorithm version we conducted an ABC cross validations analysis. That is, we selected one trajectory from the million reference set and used it as the real one. We did this selection in a random manner but with the condition that the parameters chosen were not close to the upper limit of the prior distribution. We consider $\lambda\leq25$ and $\kappa\leq 70$. Then, parameters were estimated using different threshold values ($\epsilon$) with the three algorithms and using all simulations except the chosen one. This process was replicated $N_{rep}=100$ times. For each method and $\epsilon$ value, we recorded the posterior samples obtained for both $\lambda$ and $\kappa$. We then calculated the prediction error as

\begin{equation}
\sqrt{\frac{\sum_i(\tilde{\theta_i}-\theta_i)^2}{N_{rep}}}   \hspace{2cm}     \theta=(\lambda,\kappa);  i=1,...,N_{rep}
\label{eq_error}
\end{equation}
where $\theta_i$ is the true parameter value of the $i$th simulated data set and $\tilde\theta_i$ is the posterior median of the parameter. We also compute a dispersion measure of the errors in relation to the magnitude of the parameters for each method and tolerance value. We call $MD$ to this index and we calculate it as

\begin{equation}
MD=\left[\displaystyle\sum_i\frac{|\tilde{\theta_i}-\theta_i|}{\theta_i}\right]/N_{rep}
\hspace{2cm}     \theta=(\lambda,\kappa);  i=1,...,N_{rep}
\label{eq_md}
\end{equation}

Furthermore, in order to assess whether the spread of the posterior distributions were not overly large or small, we computed the empirical coverage of the  $\alpha=95\%$ credible interval for the two parameters and for different thresholds ($\epsilon$). The empirical coverage is the proportion of simulations for which the true value of the parameter falls within the $\alpha\%$ highest posterior density interval (HPD). If the nominal confidence levels were accurate, this proportion should have been near $0.95$. 
  If this is true for all $\alpha$, it is said that the analysis satisfies the \textit{coverage property}. A way to test this property is by performing the Coverage Test and it is also a useful way to choose the threshold value $\epsilon$. 
  This test was first introduced by \citet{prangle_diagnostic_2014} and the basic idea is to perform ABC analyses under many data sets simulated from known parameter values and for each of them compute $p$, the proportion of the estimated posterior distribution smaller than the true parameter. Ideally these values should be distributed as a $U(0, 1)$.
For a complete description of this test see \citep{prangle_diagnostic_2014}.
In order to analyze all possible $\alpha$ values, we performed this test using the package ``abctools'' (\citet{nunes_abctools:_2015}; https://cran.r-project.org/web/packages/abctools/index.html). 
 
 \subsubsection*{Relative scale of observations and accuracy of the posterior density}\label{exp2}

\begin{figure}
 \centering
 \includegraphics[width=1\textwidth]{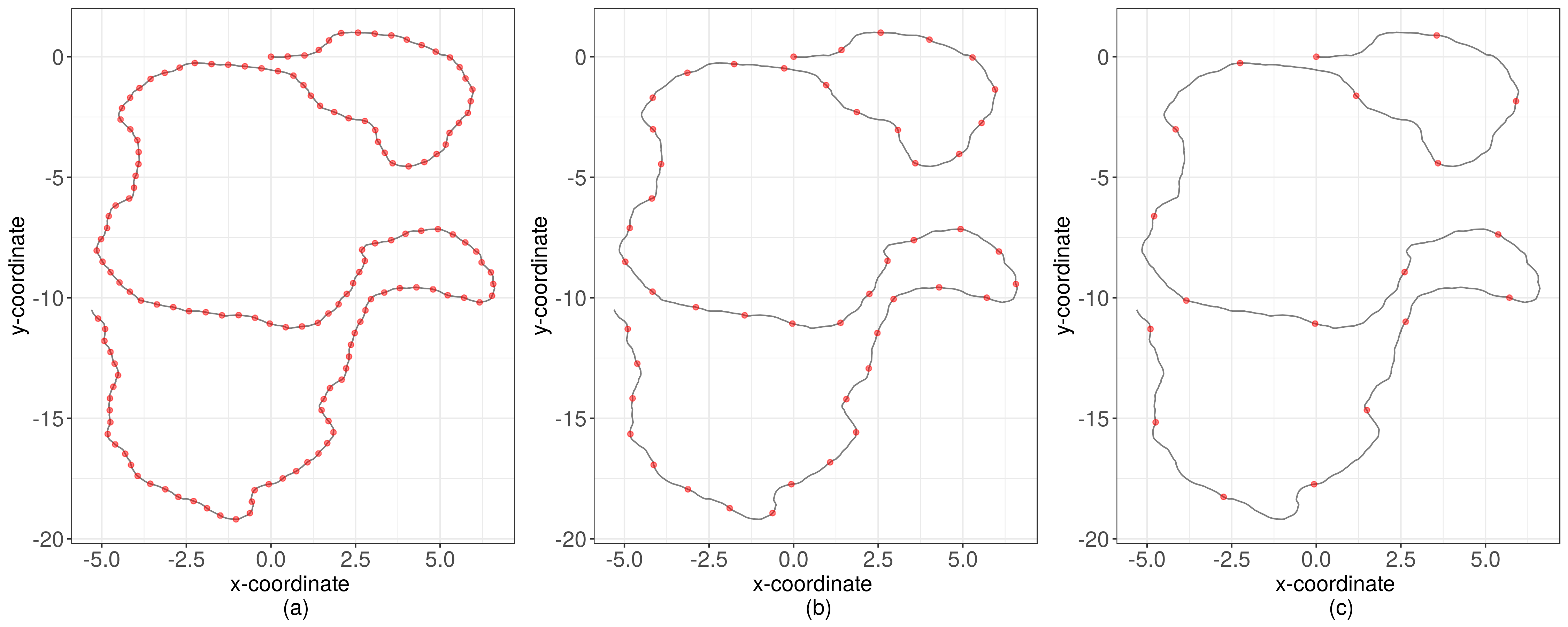}
 \caption{Simulations examples. Black lines: latent process. Red points: observation process. Different relations between the temporal scales of the observation process and the movement one: (a)over-sampling case and (c)sub-sampling case}
 \label{trayectoria}
\end{figure}

We continued the analysis evaluating how well these methods approximate posterior probabilities as a function of the ratio, $R$, between the temporal scale of observation ($\Delta t$) and the temporal scale for changes in directions ($1/\lambda$) (Figure \ref{trayectoria}). For instance, if $R=1$ then $\lambda =1/\Delta t$, which means that the time between consecutive observations is equal to the mean time between changes in direction. Conversely, if $R<1$ then the time scale between consecutive observations is smaller than the time scale at which animals decide to change directions, and the opposite occurs if $R>1$ (Figure \ref{trayectoria}).    
We considered different values of $R$ (between $0.06$ and $5$) and for each simulated $50$ trajectories with values of $\kappa \in \{10, 20, 30, \ldots, 70\}$. 
Then, using the original million set, the estimations for the three methods of ABC were computed considering these new trajectories as the true observations . We calculated the predictor error for $\kappa$ and $\lambda$ for every combination of $R$ and $\kappa$. 

\subsection*{Real data example} 
We evaluate the performance of this model using data from a real trajectory reconstructed in high-resolution using information from DailyDiary and GPS devices. 
With a high resolution trajectory it is possible to infer in such manner the moments in which the animal changes the direction and use that information to estimate the mean of the step length and the dispersion of the turning angles. In this manner it is possible to obtain certain ``true'' parameter values to then see if they can be recovered using the ABC techniques.  

The data used were collected from one sheep in Bariloche, Argentina, during February and March of 2019. The sheep was equipped with a collar containing a GPS (CatLog-B, Perthold Engineering, www.perthold.de; USA), that was programmed to record location data every five minutes, and a DailyDiary (DD, \citep{wilson_prying_2008}), that was programmed to record $40$ acceleration data per second (frequency of $40$Hz) and $13$ magnetometer data per second (frequency of $13$Hz). The DD are electronic devices that measure acceleration and magnetism in three dimensions, which can be described relative to the body of the animal. Such data allow the Dead-Reckoned (DR) path of an animal to be reconstructed at high resolution. The goal here is to use the detailed observed trajectories to decompose them in steps and turns and then test if the ABC approach can estimate the parameters of the distributions for steps and turns.

From the original data, we randomly selected one segment of $6$ hours. Using the DD information, we first estimated the path traveled (pseudotrack) by the sheep using the dead-reckoning technique. \citep{wilson_dead_1988, rory_p.wilson_all_2007}. In this step we made use of the R package ``TrackReconstruction''  (https://cran.r-project.org/web/packages/ TrackReconstruction/index.html). After that, we corrected the bias of those estimation using the data from the GPS \citep{yang_liu_bias_2015}. This correction was made using the R package ``BayesianAnimalTracker'' (https://cran.r-project.org/web/packages/ BayesianAnimalTracker/index.html). In this way, we obtained a trajectory sampled with a resolution of 1 second. To satisfy the hypotheses of the model, we selected part of that trajectory that appeared to come from the same behaviour, i.e we selected a piece of the trajectory that visually appeared to have the same distribution of turn angles and step lengths.

In order to estimate the parameters of the trajectory it was necessary to determine the points in which there was a change of movement direction. We applied the algorithm proposed by Potts. et al. \citep{potts_finding} that detects the turning points of the trajectory using data of the animal headings and subsequently calculated steps and turning angles. So, in that way we not only obtained the values of the $N_j$, but also we obtained samples for the step's length and the turning angles. With that information it was really simple to infer the parameter's values via MCMC techniques and obtain samples from the joint posterior distribution using the Stan software \citep{carpenter_stan:_2017}.

Then, we calculated the summary statistics of the trajectory observed at $dt=50$ secs ($1$ observation every $50$ of the reconstructed trajectory) and applied the three ABC algorithms. We finally compared both estimations.

\section*{Results}

\subsection*{Assessment of the inference capacity of the ABC methods}
\begin{figure}
\centering

  \includegraphics[width=0.8\textwidth]{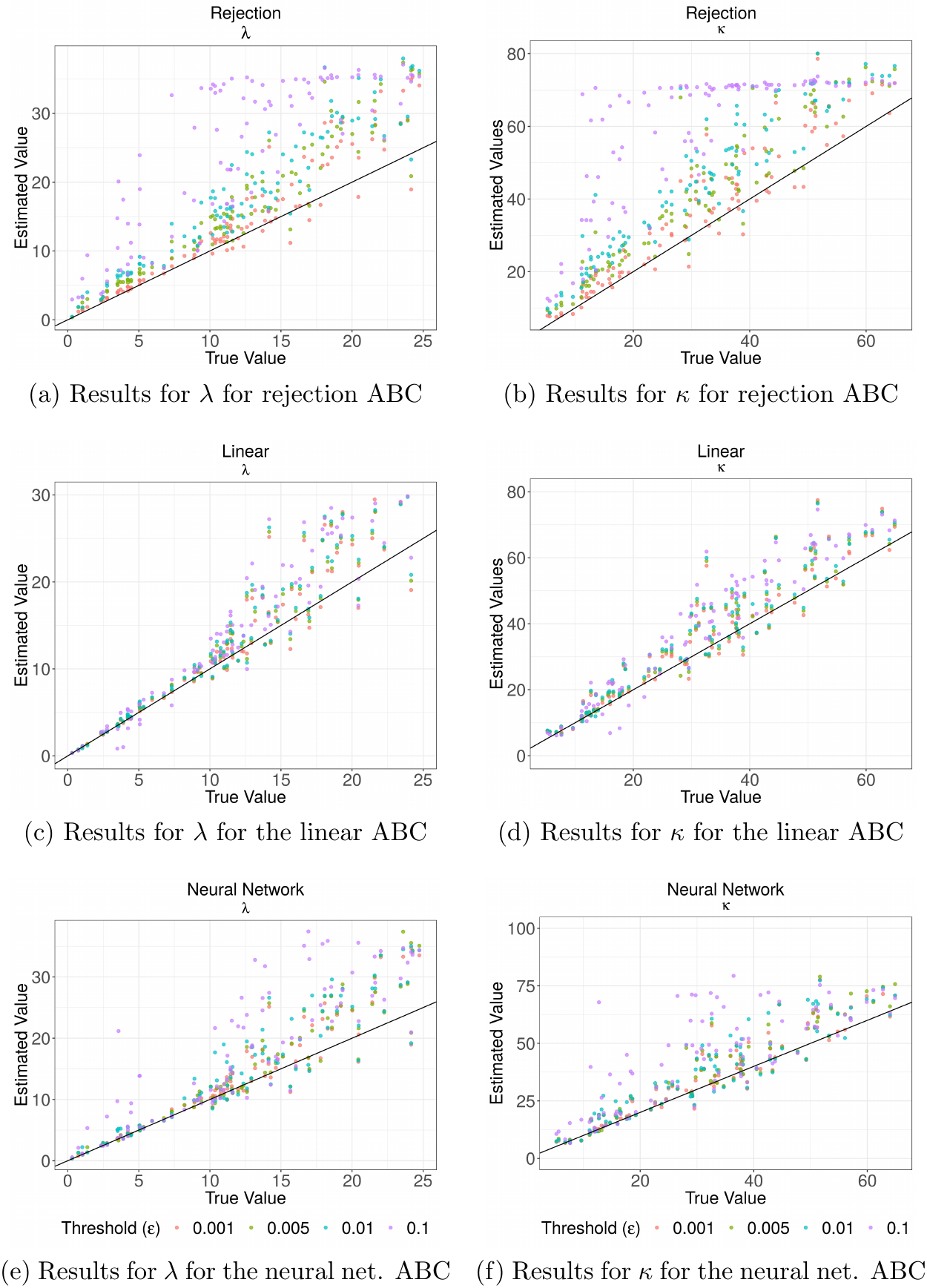}

\caption{Results for the cross validation analysis for the Rejection ABC algorithm and the two corrections. 
    Relation between the real parameters and the median of the estimated posterior are shown.
    With different colors the results for the different values of tolerance ($\epsilon$): purple for $\epsilon=0.1$, blue for $\epsilon=0.01$,
   green $\epsilon=0.005$ and red for $\epsilon=0.001$.The black line indicates the line $x = y$ - the ideal relation.}
\label{CV}
\end{figure}

Figure \ref{eq_error} shows the values of the prediction errors and the dispersion index ($MD$) for each method and $\epsilon$ value obtained from the ABC cross validation analysis. 
In all cases the prediction errors decreased when the value of the threshold ($\epsilon$) decreased . 
However, for the algorithms corrected via linear regression and neural networks larger threshold levels ($\epsilon$) can produce lower prediction error. Something similar happened with the $MD$ values: lower threshold values imply lower values of this index (Figure \ref{eq_error}). These values give us an idea of the width of the posterior distributions. It is evident that for the case of the rejection algorithm the posterior distributions are quite wide, especially for $\epsilon=0.1$. However, for the corrected algorithms we can assume that the difference between the estimated and true parameters are up to approximately $1.3$ units for $\kappa$ and in $0.3$ for $\lambda$ in the best case ($\epsilon=0.001$).  
Figure \ref{CV} shows that the estimate of $\lambda$ improves when it takes lower values, especially for the algorithm corrected via linear regression. We discuss this point in the Discussion. Based on these results, the algorithm corrected via linear regression seems to be the one that presents the best performance.

\begin{figure}[ht]
 \centering
 \includegraphics[width=1\textwidth]{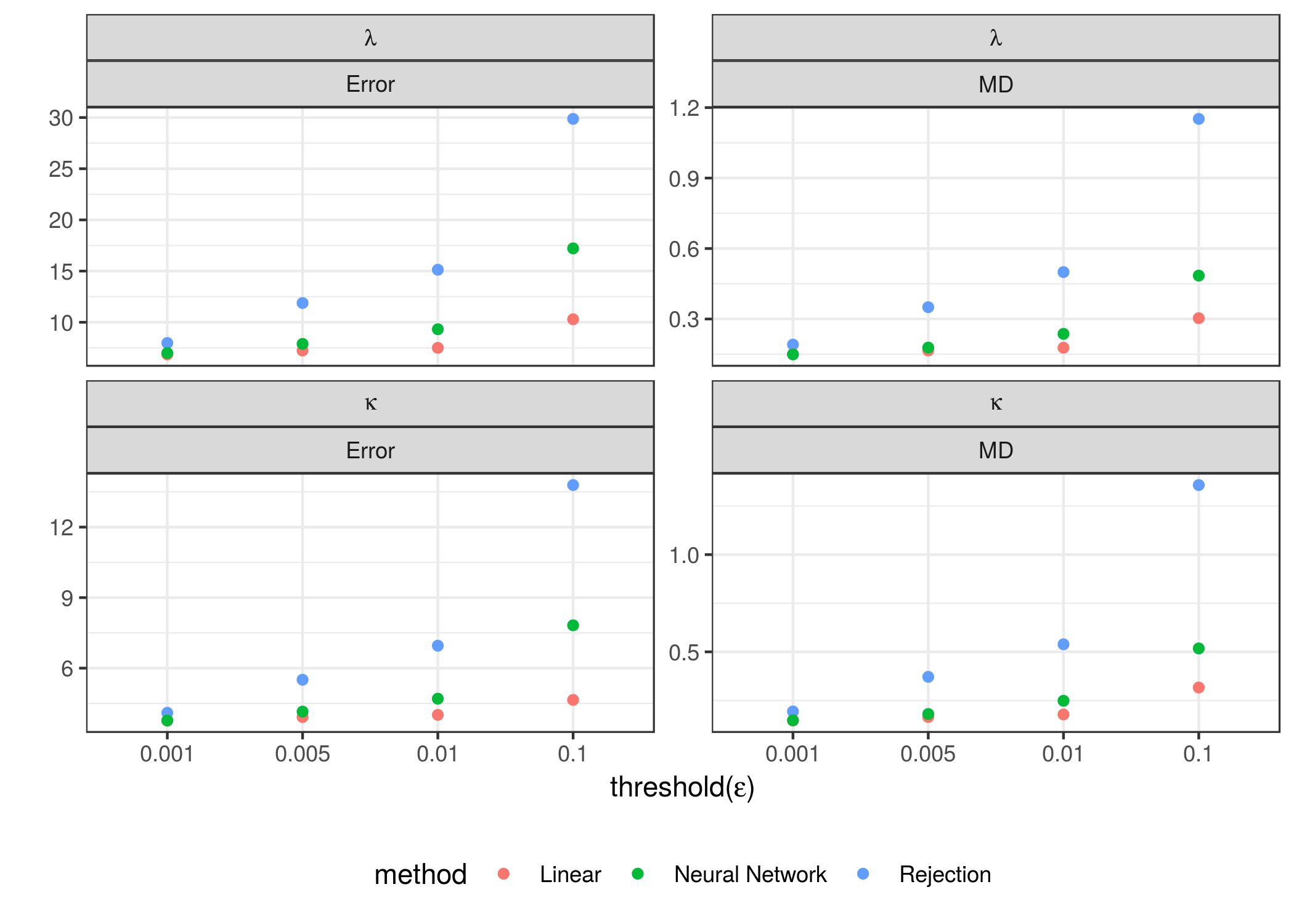}
 \caption{Values for the Prediction errors (Equation \ref{eq_error}) and for the dispersion index $MD$ (Equation \ref{eq_md}) for both parameters in each method and threshold $\epsilon$}  
  \label{table2}
\end{figure}

We estimated the empirical coverage of the $95\%$ HPD intervals for both parameters ($\kappa$ and $\lambda$), for the three ABC algorithms and for different threshold ($\epsilon$) values. Almost always these indices were greater than $95$, except in the case of the highest threshold value ($\epsilon=0.1$) for the simple rejection algorithm and the one corrected via neural network, for which the empirical coverages were a little below $0.95$. The plots for this analysis are provided in the Appendix.

\begin{figure}
    \centering
        \includegraphics[width=1\textwidth]{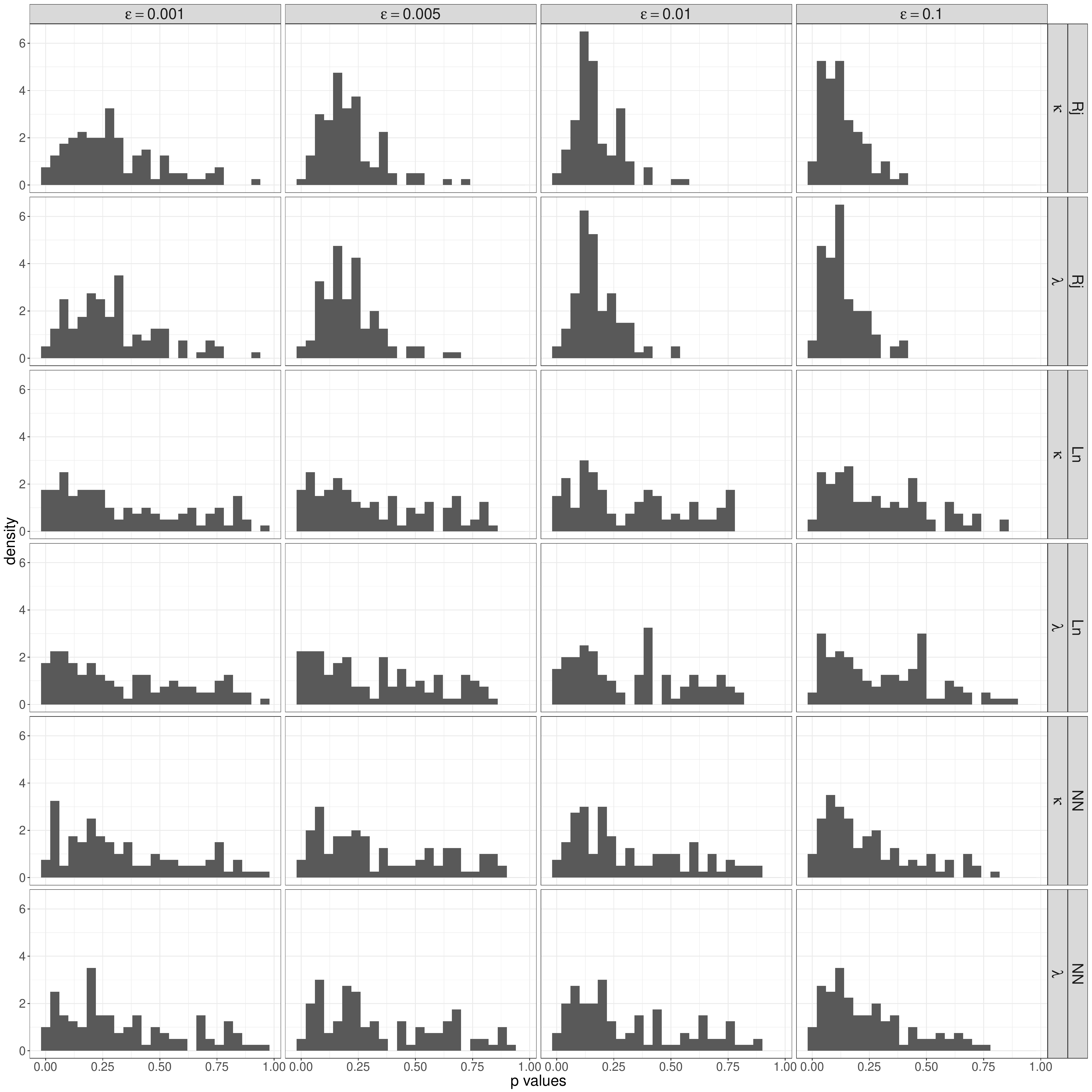}
           \caption{Coverage analyses for parameter estimation. Relative frequency $p$ of accepted parameter values that were less than the true value in ABC analyses. By column the results for different $\epsilon$ values. 
           The first two rows correspond to the results of Rejection algorithm, the second two to the ABC with linear correction, and the last two to the ABC corrected via neural networks.  }
               \label{PlotCvTest.Rate}
\end{figure}

Finally, in order to check the \textit{coverage property} we performed a coverage test for both parameters (Figure \ref{PlotCvTest.Rate}). 
In most cases, the distributions obtained do not show a clear approximation of a $U(0,1)$. However there is an evident difference between the histograms obtained with the simple rejection ABC and the histograms obtained with the other two algorithms. The  shapes  of  the rejection ABC are those that are farthest from being uniform: for both parameters the distributions of the p values are left skewed indicating that the algorithm tends to overestimate the parameters. For the other two algorithms the left skewed is much moderate, and even in the case of the lowest $\epsilon$ values for the linear algorithm the histograms are more uniform, indicating that its \textit{coverage} could be being reached. However, 
not rejecting that coverage holds does not unequivocally demonstrate 
that the ABC posterior approximation is accurate. If the empirical data is uninformative, the ABC will return posterior distributions very similar to the priors, which would produce uniform coverage plots.

\subsection*{Relative scale of observations and accuracy of the posterior density}

\begin{figure}[ht]
    \centering
        \includegraphics[width=1\textwidth]{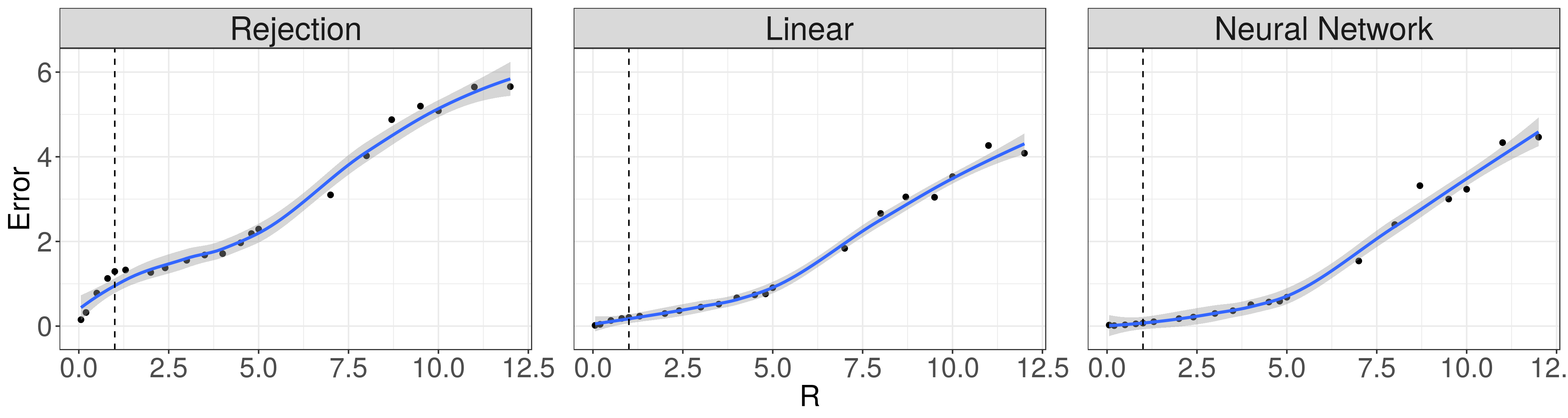}
           \caption{Scatter plot-and-smoother graph (by local regression) of the prediction errors for the expected time between changes of direction ($\lambda$) for different ratios between the temporal scale of observation and the scale for changes in directions ($R$). High $R$ values indicate that the temporal scale of the observation process is higher than the temporal scale of the the movement decision process. 
           The black dots are the prediction errors for $\lambda$ for each $R$ value. The blue line is the smoothed curve for those values. The $95\%$ intervals are shown in grey. The vertical dotted line indicates $R=1$. }
               \label{PredlambdaError.Rate}
\end{figure}

\begin{figure}
    \centering
        \includegraphics[width=1\textwidth]{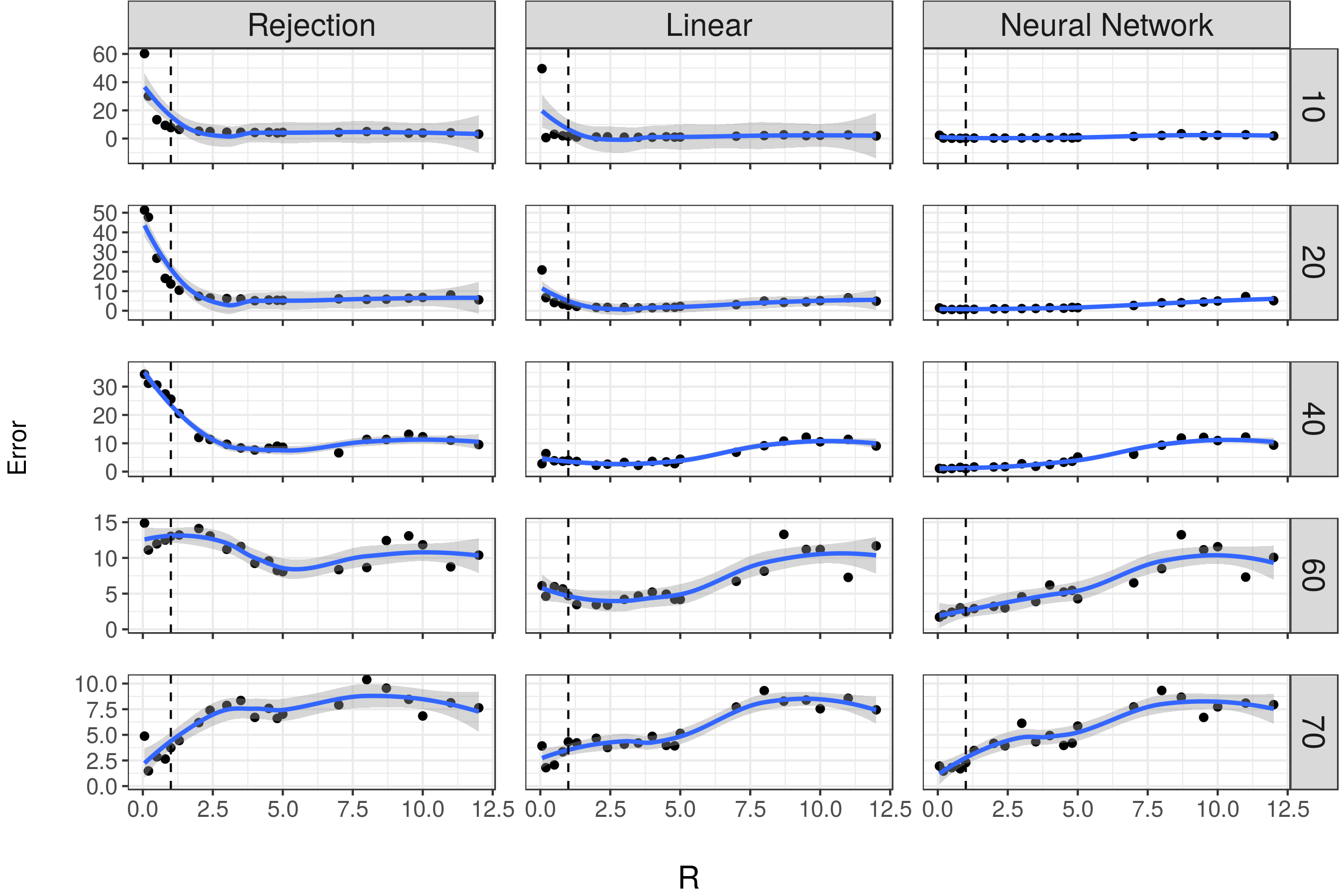}
      
         \caption{Scatter plot-and-smoother graph (by local regression) of the prediction errors for the dispersion parameter of the turning angles ($\kappa$) for different ratios between the temporal scale of observation and the scale for changes in directions ($R$). Rows correspond to different values of $\kappa$ and columns to different ABC algorithms. High $R$ values indicate that the temporal scale of the observation process is larger than the temporal scale of the the movement decision process. 
         Black dots are the prediction errors of $\kappa$ for each $R$ value. The blue line is the smoothed curve for those values. The $95\%$ intervals are shown in grey. The vertical dotted line indicates $R=1$.}
          \label{PredkappaError.Rate}
\end{figure}

In order to evaluate the importance of the relationship between the time scale of the observation process and the time in which changes occur in the latent process, we evaluated how well the two parameters fit in relation to the R ratio. The prediction errors for $\lambda$ increased as the value of $R$ increased (Figure \ref{PredlambdaError.Rate}). 
For the case of the prediction errors for $\kappa$ 
this relation can be seen when the true value of this parameter takes larger values (Figure \ref{PredkappaError.Rate}). 

For the case of the prediction errors for $\kappa$ this relation can be seen when the true value of this parameter takes larger values. Again, the corrected algorithms have the smallest errors for both parameters.


According to the above results, it is evident that there is a relationship between the ratio $R$ and the capacity of these methods to estimate the parameters. For rates approximately less than $5$ the errors are small and it is possible to obtain good estimates. This necessitates that the time scale of the observation process be approximately less than $5$ times the time-scale at which the animals decide to change direction. For higher values of $\Delta t$ it would be more difficult to make inferences using this technique. 

\subsection*{Sheep data}
\begin{figure}[H]
    \centering
        \includegraphics[width=0.5\textwidth]{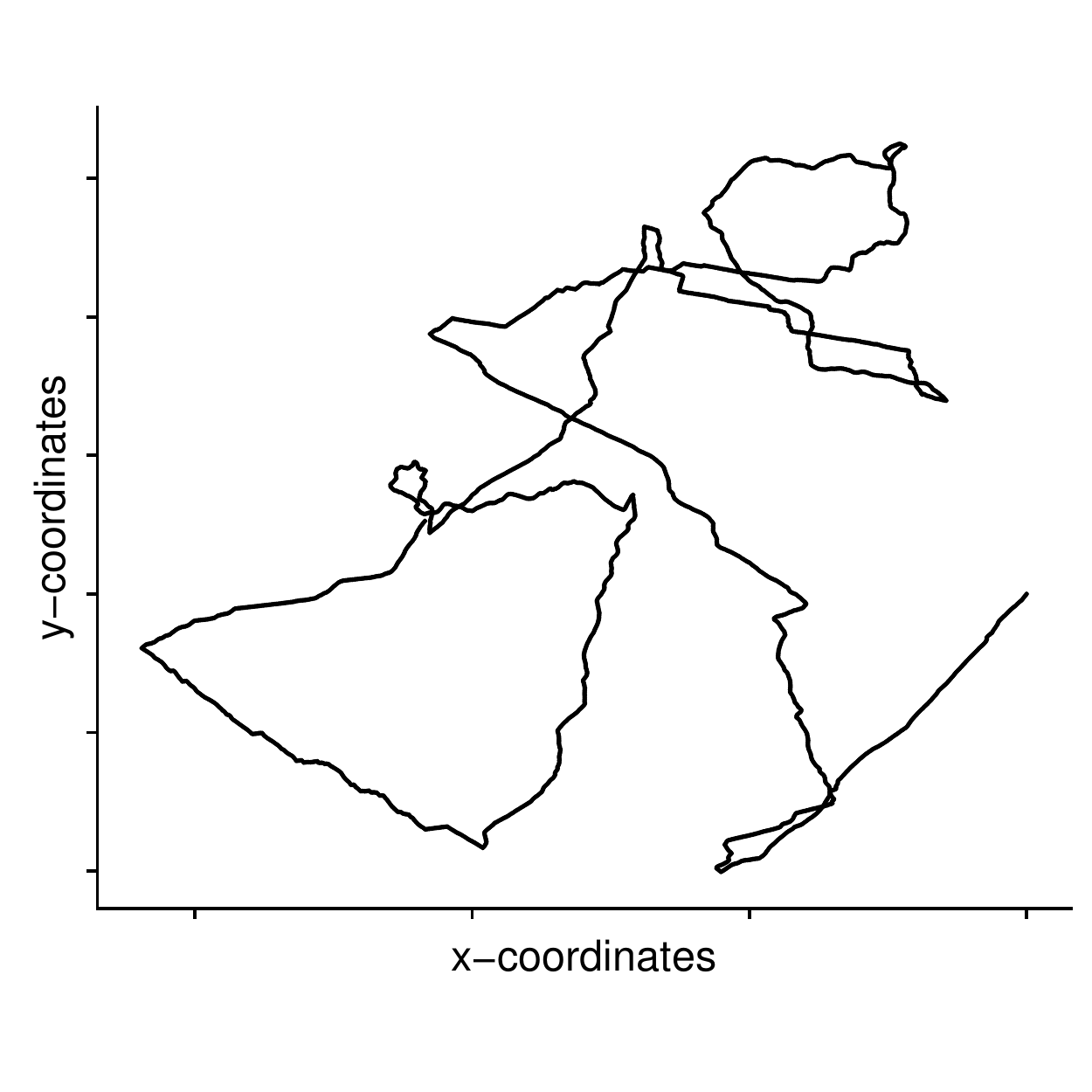}
      
         \caption{Final trajectory with 1 second resolution reconstructed by dead-reckoning and corrected using the GPS information. }
          \label{realtrajd}
\end{figure}

The selected trajectory was of February $27$, $2019$ from $19:01:21$hs to $20:02:00$hs, a total of $1.01$ hours (Figure \ref{realtrajd}). The posterior distribution of each parameter was estimated from a sample of $3\times 1000$ independent MCMC observations.
As they were in the lower limits of the prior distributions we simulate a new set of trajectories with priors of $\kappa\sim U[0,10]$ and $\lambda\sim U[0,10]$ and compute the ABC inferences with those simulations.

\begin{figure}[H]
\centering

  \includegraphics[width=0.8\textwidth]{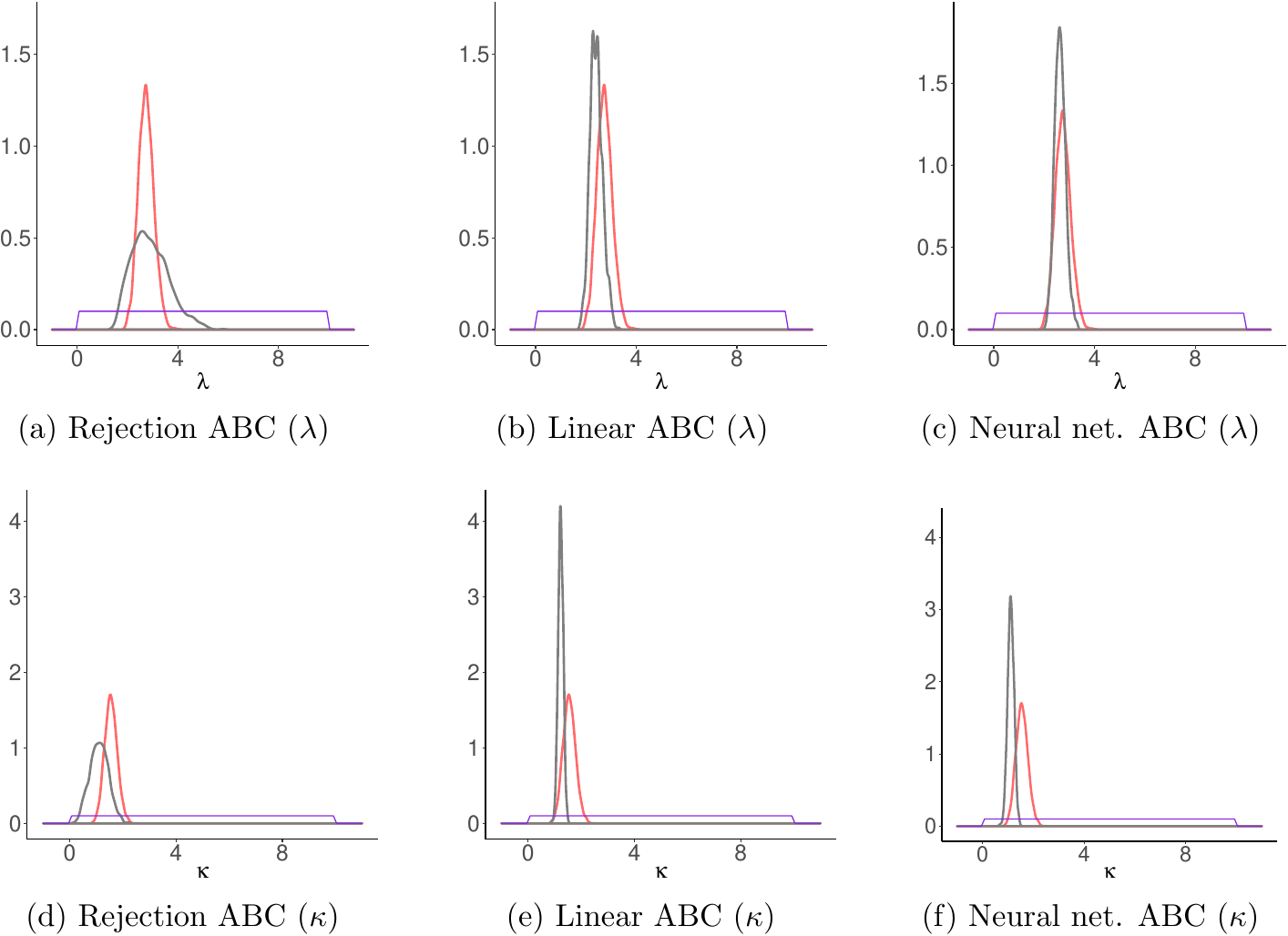}

\caption{Results for the real data examples. With red, the posteriors obtained for both parameters $\kappa$ and $\lambda$ through MCMC and with gray those obtained with each ABC algorithm. By row the results for each parameters and by column the results for each ABC algorithm.}

\label{real_plots}
\end{figure}

The posteriors obtained through MCMC and through the ABC algorithms gave similar results (\ref{real_plots}). Again, the rejection ABC algorithm produced the estimation which is less exact, i.e the posterior is the furthest from the one obtained by MCMC. 
 Although this trajectory is just a simple example, it shows that it is possible to apply this model to actual animal trajectories.

\section*{Discussion}


Animal movement modelling and analysis is either considered in continuous or discrete time. Continuous time models are more realistic but often harder to interpret than the discrete versions \citep{mcclintock_when_2014}. A compromise between these approaches is to model movement as steps and turns but to have step duration (or the times at which turns are made) occurring in continuous time. In this way one could get the best of both worlds, so to speak. Here we considered that the underlying movement process, evolving in continuous time, is observed at regular time intervals as would be standard for a terrestrial animal fitted with a GPS collar. 
The likelihood function resulted to be complex to calculate, but it is feasible to quickly generate simulations from the process and observation models. Thus, we proposed to use ABC methods.  Even though these techniques showed certain limitations, it was possible to obtain accurate parameter estimates when the temporal scale of observations was not too coarse compared to the scale of changes in direction. 

Our simulation study showed that simple rejection ABC does not perform well for the proposed state-space model but the two corrected version of this algorithm really improve the estimations (Table \ref{table2} and Figure \ref{CV}). Overall, the best performance was obtained with the linear correction. However, the applicability of these methods depends strongly on the rate between the observation process’s scale and the mean time between changes in movement direction. We found that when this ratio is smaller than $5$ it is possible to make inferences about the parameters (Figures  \ref{PredlambdaError.Rate} and \ref{PredkappaError.Rate}). That is, it would be necessary that the observations are less than $5$ times the average of the times between changes of directions in order to be able to generate good estimations.    

Beyond our findings about the capacity to make inference with these techniques in a simulation study, it is important to note that in an applied case more informative priors could be considered. Here, our aim was to evaluate the performance of the ABC techniques considering several parameter combinations generating trajectories and then sampling from those trajectories. In order to optimize computing time, we simulated a million trajectories sampling their parameters from uniform distributions and then we randomly choose one of them as observed data while the rest of the simulations was used to perform the ABC computations. That justify the use of uniform priors for our parameters. 
 As we did in our real data example, in applied cases it wold be relatively straightforward to come up with more informative priors, especially for the expected time for changes in movement direction.   

The movement model presented here is quite simple as we assume constant movement speed and turning angles with zero mean. Nevertheless, the model is an improvement over discrete-time versions where the temporal scale of movement has to match the scale of observations. Further developments of these methods should consider additional features that are common in movement studies such as including more than one movement behavior and the effect of habitat features on both movement parameters and changes between behaviors \citep{morales_extracting_2004, mevin_b._hooten_animal_2017}. Even though this means estimating several parameters, such models will imply further structure in the trajectories that could be used as part of the summary statistics used to characterize the data.  Hence, it might reduce the combination of parameters values capable of reproducing features present in the data, allowing for ABC inference.

As new technologies allow us to obtain very detailed movement data, we can have better estimates of the temporal scales at which animals make movement decisions. As we did in our real data example,  high-frequency data from accelerometers and magnetometers combined with GPS data can be used to obtain trajectories with sub-second temporal resolution to then detect fine-scale movement decisions such as changes in direction. 
These detailed trajectories could be used to elicit informative priors to use when only coarser data is available.

In general, the processes behind the realized movement of an individual and the processes that affect how we record the trajectory are usually operating at different time scales, making it challenging to analyze and to understand the former given the latter. The state-space model used here allowed us to connect these two scales in an intuitive and easy to interpret way. Our findings underscore the idea that the time scale at which animal movement decisions are made needs to be considered when designing data collection protocols, and that high-frequency data may not be necessary to have good estimates of certain movement processes.

\section*{Acknowledgements}
The authors are grateful to Agustina Di Virgilio, Teresa Moran Lopez and Florencia Tiribelli for their useful comments that substantially improved the manuscript.

\section*{Authors' Contributions}
JMM conceived the ideas; All the authors designed methodology; SRS simulated the data and performed the analysis; SRS and JMM led the writing of the manuscript.  All authors contributed critically to the drafts and gave final approval for publication.

\bibliography{Mibiblioteca.bib}

\begin{thebibliography}{42}
\providecommand{\natexlab}[1]{#1}
\providecommand{\url}[1]{\texttt{#1}}
\expandafter\ifx\csname urlstyle\endcsname\relax
  \providecommand{\doi}[1]{doi: #1}\else
  \providecommand{\doi}{doi: \begingroup \urlstyle{rm}\Url}\fi

\bibitem[Baudet et~al.(2015)Baudet, Donati, Sinaimeri, Crescenzi, Gautier,
  Matias, and Sagot]{baudet_cophylogeny_2015}
C.~Baudet, B.~Donati, B.~Sinaimeri, P.~Crescenzi, C.~Gautier, C.~Matias, and
  M.-F. Sagot.
\newblock Cophylogeny reconstruction via an approximate {Bayesian} computation.
\newblock \emph{Syst. Biol.}, 64\penalty0 (3):\penalty0 416--431, 2015.

\bibitem[Beaumont(2010)]{beaumont_approximate_2010}
M.~A. Beaumont.
\newblock Approximate {Bayesian} {Computation} in {Evolution} and {Ecology}.
\newblock \emph{Annual Review of Ecology, Evolution, and Systematics},
  41:\penalty0 379--406, 2010.

\bibitem[Beaumont et~al.(2002)Beaumont, Zhang, and
  Balding]{beaumont_approximate_2002}
M.~A. Beaumont, W.~Zhang, and D.~J. Balding.
\newblock Approximate {Bayesian} {Computation} in {Population} {Genetics}.
\newblock \emph{Genetics}, 162:\penalty0 2025--2035, 2002.

\bibitem[Bertorelle et~al.(2010)Bertorelle, Benazzo, and
  Mona]{bertorelle_abc_2010}
G.~Bertorelle, A.~Benazzo, and S.~Mona.
\newblock {ABC} as a flexible framework to estimate demography over space and
  time: some cons, many pros.
\newblock \emph{Molecular Ecology}, 19:\penalty0 2609--2625, 2010.

\bibitem[Blackwell(1999)]{blackwell_random_1999}
P.~Blackwell.
\newblock Random diffusion models for animal movement.
\newblock \emph{Ecological Modelling}, 100:\penalty0 87--102, 1999.

\bibitem[Blum and François(2010)]{blum_non-linear_2010}
M.~G.~B. Blum and O.~François.
\newblock Non-linear regression models for {Approximate} {Bayesian}
  {Computation}.
\newblock \emph{Stat Comput}, 20:\penalty0 63--73, 2010.

\bibitem[Carpenter et~al.(2017)Carpenter, Gelman, Hoffman, Lee, Goodrich,
  Betancourt, Brubaker, Guo, Li, and Riddell]{carpenter_stan:_2017}
B.~Carpenter, A.~Gelman, M.~D. Hoffman, D.~Lee, B.~Goodrich, M.~Betancourt,
  M.~Brubaker, J.~Guo, P.~Li, and A.~Riddell.
\newblock Stan: {A} {Probabilistic} {Programming} {Language}.
\newblock \emph{Journal of Statistical Software}, 76\penalty0 (1):\penalty0
  1--32, Jan. 2017.
\newblock ISSN 1548-7660.
\newblock \doi{10.18637/jss.v076.i01}.

\bibitem[Csill\'ery et~al.(2010)Csill\'ery, Blum, Gaggiotti, and
  François]{csillery_approximate_2010}
K.~Csill\'ery, M.~G.~B. Blum, O.~E. Gaggiotti, and O.~François.
\newblock Approximate {Bayesian} {Computation} ({ABC}) in practice.
\newblock \emph{Trends in Ecology \& Evolution}, 25:\penalty0 410--418, 2010.

\bibitem[Csill\'ery et~al.(2012)Csill\'ery, François, and
  Blum]{csillery_abc:_2012}
K.~Csill\'ery, O.~François, and M.~G.~B. Blum.
\newblock abc: an {R} package for approximate {Bayesian} computation ({ABC}).
\newblock \emph{Methods in Ecology and Evolution}, 3:\penalty0 475--479, 2012.

\bibitem[Fèvre et~al.(2006)Fèvre, Bronsvoort, Hamilton, and
  Cleaveland]{fevre_animal_2006}
E.~M. Fèvre, B.~M. d.~C. Bronsvoort, K.~A. Hamilton, and S.~Cleaveland.
\newblock Animal movements and the spread of infectious diseases.
\newblock \emph{Trends Microbiol.}, 14:\penalty0 125--131, 2006.

\bibitem[Gurarie and Ovaskainen(2011)]{gurarie_characteristic_2011}
E.~Gurarie and O.~Ovaskainen.
\newblock Characteristic {Spatial} and {Temporal} {Scales} {Unify} {Models} of
  {Animal} {Movement}.
\newblock \emph{The American Naturalist}, 178\penalty0 (1):\penalty0 113--123,
  July 2011.

\bibitem[Harris and Blackwell(2013)]{harris_flexible_2013}
K.~J. Harris and P.~G. Blackwell.
\newblock Flexible continuous-time modelling for heterogeneous animal movement.
\newblock \emph{Ecological Modelling}, 255:\penalty0 29--37, 2013.

\bibitem[Johnson et~al.(2008)Johnson, London, Lea, and
  Durban]{johnson_continuous-time_2008}
D.~S. Johnson, J.~M. London, M.-A. Lea, and J.~W. Durban.
\newblock Continuous-time correlated random walk model for animal telemetry
  data.
\newblock \emph{Ecology}, 89:\penalty0 1208--1215, 2008.

\bibitem[Jonsen et~al.(2005)Jonsen, Flemming, and Myers]{jonsen_robust_2005}
I.~D. Jonsen, J.~M. Flemming, and R.~A. Myers.
\newblock Robust {State-}{Space} {Modeling} of {Animal} {Movement} {Data}.
\newblock \emph{Ecology}, 86:\penalty0 2874--2880, 2005.

\bibitem[{Kurt Hornik} and {Bettina Gr\"{u}n}(2014)]{kurt_hornik_maximum_2014}
{Kurt Hornik} and {Bettina Gr\"{u}n}.
\newblock On maximum likelihood estimation of the concentration parameter of
  von {Mises}–{Fisher} distributions.
\newblock \emph{Comput Stat}, 29\penalty0 (5):\penalty0 945--957, 2014.

\bibitem[{Liu, Y.} et~al.(2015){Liu, Y.}, {Battaile, B.C.}, {Trite, A W.}, and
  {Zidek, J.V.}]{yang_liu_bias_2015}
{Liu, Y.}, {Battaile, B.C.}, {Trite, A W.}, and {Zidek, J.V.}
\newblock Bias correction and uncertainty characterization of {Dead}-{Reckoned}
  paths of marine mammals.
\newblock \emph{Animal Biotelemetry}, 3:\penalty0 51, 2015.

\bibitem[Lopes and Beaumont(2010)]{lopes_abc:_2010}
J.~S. Lopes and M.~A. Beaumont.
\newblock {ABC}: a useful {Bayesian} tool for the analysis of population data.
\newblock \emph{Infect. Genet. Evol.}, 10:\penalty0 826--833, 2010.

\bibitem[Marjoram and Tavaré(2006)]{marjoram_modern_2006}
P.~Marjoram and S.~Tavaré.
\newblock Modern computational approaches for analysing molecular genetic
  variation data.
\newblock \emph{Nat. Rev. Genet.}, 7\penalty0 (10):\penalty0 759--770, 2006.

\bibitem[Marjoram et~al.(2003)Marjoram, Molitor, Plagnol, and
  Tavaré]{marjoram_markov_2003}
P.~Marjoram, J.~Molitor, V.~Plagnol, and S.~Tavaré.
\newblock Markov chain {Monte} {Carlo} without likelihoods.
\newblock \emph{PNAS}, 100:\penalty0 15324--15328, 2003.

\bibitem[Matthiopoulos et~al.(2015)Matthiopoulos, Fieberg, Aarts, Beyer,
  Morales, and Haydon]{matthiopoulos_establishing_2015}
J.~Matthiopoulos, J.~Fieberg, G.~Aarts, H.~L. Beyer, J.~M. Morales, and D.~T.
  Haydon.
\newblock Establishing the link between habitat selection and animal population
  dynamics.
\newblock \emph{Ecological Monographs}, 85:\penalty0 413--436, 2015.

\bibitem[McClintock et~al.(2012)McClintock, King, Thomas, Matthiopoulos,
  McConnell, and Morales]{mcclintock_general_2012}
B.~T. McClintock, R.~King, L.~Thomas, J.~Matthiopoulos, B.~J. McConnell, and
  J.~M. Morales.
\newblock A general discrete-time modeling framework for animal movement using
  multistate random walks.
\newblock \emph{Ecological Monographs}, 82:\penalty0 335--349, 2012.

\bibitem[McClintock et~al.(2014)McClintock, Johnson, Hooten, Ver~Hoef, and
  Morales]{mcclintock_when_2014}
B.~T. McClintock, D.~S. Johnson, M.~B. Hooten, J.~M. Ver~Hoef, and J.~M.
  Morales.
\newblock When to be discrete: the importance of time formulation in
  understanding animal movement.
\newblock \emph{Movement Ecology}, 2:\penalty0 21, 2014.

\bibitem[McKinley et~al.(2009)McKinley, Cook, and
  Deardon]{mckinley_inference_2009}
T.~McKinley, A.~R. Cook, and R.~Deardon.
\newblock Inference in {Epidemic} {Models} without {Likelihoods}.
\newblock \emph{The International Journal of Biostatistics}, 5\penalty0 (1),
  2009.

\bibitem[{Mevin B. Hooten} et~al.(2017){Mevin B. Hooten}, {Devin S. Johnson},
  {Brett T. McClintock}, and {Juan M. Morales}]{mevin_b._hooten_animal_2017}
{Mevin B. Hooten}, {Devin S. Johnson}, {Brett T. McClintock}, and {Juan M.
  Morales}.
\newblock \emph{Animal {Movement}: {Statistical} {Models} for {Telemetry}
  {Data}.}
\newblock CRC Press, 2017.

\bibitem[Morales et~al.(2004)Morales, Haydon, Frair, Holsinger, and
  Fryxell]{morales_extracting_2004}
J.~M. Morales, D.~T. Haydon, J.~Frair, K.~E. Holsinger, and J.~M. Fryxell.
\newblock Extracting {More} {Out} of {Relocation} {Data}: {Building} {Movement}
  {Models} as {Mixtures} of {Random} {Walks}.
\newblock \emph{Ecology}, 85:\penalty0 2436--2445, 2004.

\bibitem[Morales et~al.(2010)Morales, Moorcroft, Matthiopoulos, Frair, Kie,
  Powell, Merrill, and Haydon]{morales_building_2010}
J.~M. Morales, P.~R. Moorcroft, J.~Matthiopoulos, J.~L. Frair, J.~G. Kie, R.~A.
  Powell, E.~H. Merrill, and D.~T. Haydon.
\newblock Building the bridge between animal movement and population dynamics.
\newblock \emph{Philosophical Transactions of the Royal Society of London B:
  Biological Sciences}, 365\penalty0 (1550):\penalty0 2289--2301, July 2010.

\bibitem[Nathan et~al.(2008)Nathan, Getz, Revilla, Holyoak, Kadmon, Saltz, and
  Smouse]{nathan_movement_2008}
R.~Nathan, W.~M. Getz, E.~Revilla, M.~Holyoak, R.~Kadmon, D.~Saltz, and P.~E.
  Smouse.
\newblock A movement ecology paradigm for unifying organismal movement
  research.
\newblock \emph{PNAS}, 105:\penalty0 19052--19059, 2008.

\bibitem[Nunes and Prangle(2015)]{nunes_abctools:_2015}
M.~A. Nunes and D.~Prangle.
\newblock abctools: {An} {R} {Package} for {Tuning} {Approximate} {Bayesian}
  {Computation} {Analyses}.
\newblock \emph{The R Journal}, 7:\penalty0 17, 2015.

\bibitem[Othmer et~al.(1988)Othmer, Dunbar, and Alt]{othmer1988models}
H.~G. Othmer, S.~R. Dunbar, and W.~Alt.
\newblock Models of dispersal in biological systems.
\newblock \emph{Journal of mathematical biology}, 26\penalty0 (3):\penalty0
  263--298, 1988.

\bibitem[Patterson et~al.(2008)Patterson, Thomas, Wilcox, Ovaskainen, and
  Matthiopoulos]{patterson_state-space_2008}
T.~A. Patterson, L.~Thomas, C.~Wilcox, O.~Ovaskainen, and J.~Matthiopoulos.
\newblock State-space models of individual animal movement.
\newblock \emph{Trends Ecol. Evol. (Amst.)}, 23:\penalty0 87--94, 2008.

\bibitem[Potts et~al.(2018)Potts, Börger, Scantlebury, Bennett, Alagaili, and
  Wilson]{potts_finding}
J.~R. Potts, L.~Börger, D.~M. Scantlebury, N.~C. Bennett, A.~Alagaili, and
  R.~P. Wilson.
\newblock Finding turning‐points in ultra‐high‐resolution animal movement
  data.
\newblock \emph{Methods in Ecology and Evolution}, 2018.

\bibitem[Prangle et~al.(2014)Prangle, Blum, Popovic, and
  Sisson]{prangle_diagnostic_2014}
D.~Prangle, M.~G.~B. Blum, G.~Popovic, and S.~A. Sisson.
\newblock Diagnostic tools for approximate {Bayesian} computation using the
  coverage property.
\newblock \emph{Australian \& New Zealand Journal of Statistics}, 56\penalty0
  (4):\penalty0 309--329, Dec. 2014.

\bibitem[Pritchard et~al.(1999)Pritchard, Seielstad, Perez-Lezaun, and
  Feldman]{pritchard_population_1999}
J.~K. Pritchard, M.~T. Seielstad, A.~Perez-Lezaun, and M.~W. Feldman.
\newblock Population growth of human {Y} chromosomes: a study of {Y} chromosome
  microsatellites.
\newblock \emph{Mol. Biol. Evol.}, 16:\penalty0 1791--1798, 1999.

\bibitem[Scott. A~Sisson and Beaumont(2018)]{handbook_abc}
Y.~F. Scott. A~Sisson and M.~A. Beaumont.
\newblock Handbook of {Approximate} {Bayesian} {Computation}, 2018.

\bibitem[Sirén et~al.(2018)Sirén, Lens, Cousseau, and
  Ovaskainen]{siren_assessing_2018}
J.~Sirén, L.~Lens, L.~Cousseau, and O.~Ovaskainen.
\newblock Assessing the dynamics of natural populations by fitting
  individual-based models with approximate {Bayesian} computation.
\newblock \emph{Methods in Ecology and Evolution}, 9\penalty0 (5):\penalty0
  1286--1295, 2018.
\newblock ISSN 2041-210X.

\bibitem[Sisson et~al.(2018)Sisson, Fan, and Beaumont]{Scott2018}
S.~Sisson, Y.~Fan, and M.~Beaumont.
\newblock \emph{Handbook of{Approximate} {Bayesian} {Computation}}.
\newblock 2018.

\bibitem[Tanaka et~al.(2006)Tanaka, Francis, Luciani, and
  Sisson]{tanaka_using_2006}
M.~M. Tanaka, A.~R. Francis, F.~Luciani, and S.~A. Sisson.
\newblock Using {Approximate} {Bayesian} {Computation} to {Estimate}
  {Tuberculosis} {Transmission} {Parameters} {From} {Genotype} {Data}.
\newblock \emph{Genetics}, 173\penalty0 (3):\penalty0 1511--1520, 2006.

\bibitem[Tavaré et~al.(1997)Tavaré, Balding, Griffiths, and
  Donnelly]{tavare_inferring_1997}
S.~Tavaré, D.~J. Balding, R.~C. Griffiths, and P.~Donnelly.
\newblock Inferring {Coalescence} {Times} {From} {DNA} {Sequence} {Data}.
\newblock \emph{Genetics}, 145:\penalty0 505--518, 1997.

\bibitem[Turchin(1998)]{turchin_quantitative_1998}
P.~Turchin.
\newblock \emph{Quantitative {Analysis} of {Movement}: {Measuring} and
  {Modeling} {Population} {Redistribution} in {Animals} and {Plants}}.
\newblock Sinauer Associates, Sunderland, Massachusetts, USA., 1998.

\bibitem[Wilson et~al.(2007)Wilson, Liebsch, Davies, Quintana, Weimerskirch,
  Storch, Lucke, Siebert, Zankl, Müller, Zimmer, Scolaro, Campagna, Plötz,
  Bornemann, Teilmann, and McMahon]{rory_p.wilson_all_2007}
J.~Wilson, N.~Liebsch, I.~M. Davies, F.~Quintana, H.~Weimerskirch, S.~Storch,
  K.~Lucke, U.~Siebert, S.~Zankl, G.~Müller, I.~Zimmer, A.~Scolaro,
  C.~Campagna, J.~Plötz, H.~Bornemann, J.~Teilmann, and C.~R. McMahon.
\newblock All at sea with animal tracks; methodological and analytical
  solutions for the resolution of movement.
\newblock \emph{ScienceDirect}, 54:\penalty0 193--210, 2007.

\bibitem[Wilson and Wilson(1988)]{wilson_dead_1988}
R.~Wilson and M.~P. Wilson.
\newblock Dead reckoning a new technique for determining penguin movements at
  sea.
\newblock \emph{Meeresforschung}, 32(2):\penalty0 155--158, 1988.

\bibitem[Wilson et~al.(2008)Wilson, Shepard, and Liebsch]{wilson_prying_2008}
R.~Wilson, E.~Shepard, and N.~Liebsch.
\newblock Prying into the intimate details of animal lives: use of a daily
  diary on animals.
\newblock \emph{Endangered Species Research}, 4:\penalty0 123--137, 2008.

\end{thebibliography}

\newpage

\section*{Supporting Information}
\subsection*{Calculation of the complete data likelihood}

Lets consider the variable $M_i=(\mu_{i,1},\mu_{i,2})$ for describing the position (in $x-y$ coordinates) of the latent process by step $i$ and the variable $O_j=(o_{j,1},o_{j,2})$ for the position of the observation $j$.
Lets remember that we defined $N_j$ as the amount of steps (or changes of direction) that the animal took from time $1$ to time $j(\Delta t)$. 

We have that $\mu_{0,1}=0$ and $\mu_{0,2}=0$

For $i=1,...,N_{steps}$
\begin{center}
$\mu_{i,1}=\mu_{i-1,1}+cos(\phi_{i-1})t_{i-1}$

$\mu_{i,2}=\mu_{i-1,2}+sin(\phi_{i-1})t_{i-1}$
    
\end{center}

And then it is possible to parameterize the observation process as

$o_{0,1}=0$ and $o_{0,2}=0$ and for $j=1,...,N_{obs}$

\begin{center}
$\displaystyle o_{j,1}= \mu_{N_{j},1}+cos(\phi_{N_{j}})\left(j\Delta t
-\sum_{k<N_j-1} t_k  \right)$
    
\    

$\displaystyle o_{j,2}= \mu_{N_{j},2}+sin(\phi_{N_{j}})\left(j\Delta t
-\sum_{k<N_j-1} t_k  \right)$

\end{center}

So $o_j$ is a function of all positions $M_i$ from $i=0$ to $i=N_j$. Then $O_j=h(M_{0:N_{j}})$, where $M_{0:D}=(M_0,M_1,M_2,...,M_D)$.
Lets suppose that we know the number of changes of direction that the animal took between consecutive observations, so we know the $N_j \forall j$. Then the likelihood of the SSM with different temporal scales for an individual trajectory is given as

\begin{linenomath*}
\begin{equation*}
\begin{split}
L(\kappa,\lambda, M, O ) & = P\left(O_0=o_0,O_1=o_1,\cdots,O_{N_{obs}}=o_{N_{obs}}\right)\\
& = P\left(h(M_{0:N_1})=o_1,h(M_{0:N_2})=o_2,...,
h(M_{0:N_{N_{obs}}})=o_{N_{obs}} \right)
\end{split}
\end{equation*}
\end{linenomath*}

As $O_j=h(M_{0:N_{j}})$, in order to get a formulation of $L$ it is necessary to obtain the distributions of $M_{i}$ $(1)$ and for $O_j=h(M_{0:N_{j}})$ $(2)$.

\subsubsection*{Step 1: Formulation of $(1)$}

We are looking for a formulation for $M_i=(\mu_{i,1},\mu_{i,2})$.
We are going to consider just the variable corresponding to the x-coordinate ($\mu_{i,1}$), the second is analogous. 

We have that 
$$\mu_{i,1}=\mu_{i-1,1}+cos(\phi_{i-1})t_{i-1}$$ 
with $\phi_i \sim von Mises(\phi_{i-1},\kappa)$ and $t_i\sim Exp(\lambda)$. To obtain the distribution of $\mu_i|\mu_{i-1}$ it is necessary to obtain the distribution form of $Z=cos(\phi)t$. Using the Change of variable Theorem it is possible to calculate this distribution. To do that, lets first consider $V=g(\phi)=cos(\phi)$. We want to obtain an expression for $f_V$. Splitting the domain of $g$ and applying the Transformation Method Theorem, is obtain: 

$$\displaystyle f_V= \left( f_{\phi}(-acos(v))+f_{\phi}(acos(v))\right)\frac{1}{\sqrt{1-v^2}}I_{-1\leq v\leq 1}(v)$$

Now we can calculate $f_Z$ as $f_Z=Vt$. Again making use of the Transformation Method Theorem and using the fact that 
the times and angles are independent, it is possible to obtain the following expression 

$$\displaystyle f_{Z}(z_1)=\int f_V(\frac{z_1}{z_2})f_t(z_2) I_{\{-z_2\leq z_1\leq z_2\}}(z_1)I_{\{z_2> 0\}}(z_2) \cdot dz_2$$

Having $f_Z$ obtaining $p(\mu_i|\mu_{i-1})$ is immediate. 

\subsubsection*{Step 2: Formulation of $(2)$}
Now, we are looking for a formulation for $O_j=h(M_{0:N_j})$. Les write
\begin{center}
$O_j=(o_{j,1},o_{j,2})=\left(h_1(\mu_{0:N_j,1}),h_2(\mu_{0:N_j,2})\right)$    
\end{center}

Again, we are going to consider just the variable corresponding to the x-coordinate ($o_{j,1}$), the second is analogous.

We have that

\begin{linenomath*}
\begin{equation*}
\begin{split}
\displaystyle o_{j,1} & = 
h(\mu_{0:N_j,1})\\
& =\mu_{N_j,1}+\cos(\phi_{N_{j}})\left(j\Delta t-\sum_{k<N_j-1}t_k\right)\\
& = \mu_{N_{j},1}+V_{N_j}(c_j-W_{N_{j}-1})
\end{split}
\end{equation*}
\end{linenomath*}

We already know the distribution of $V_{N_j}$. The distribution of $W_{N_{j}-1}$ is just a sum of $N_{j}-2$ $Exp(\lambda)$, a $\Gamma(N_{j}-2,\lambda)$. If we consider $\tilde W_{N_{j}-1}= c_j-W_{N_{j}-1}$ (which differs with $W_{N_{j}-1}$ just in a constant), we have that $f_{\tilde W_{N_{j}-1}} (v)= f_{W_{N_{j}-1}}(c_j-v)$    

So, we can rewrite $o_{j,1}$ as

$$ o_{j,1}=\mu_{N_{j},1}+(V_{N_j})(\tilde W_{N_{j}-1})=\mu_{N_j,1}+S_{N_{j}-1}$$

To obtain the distribution of $S_{N_{j}-1}$, again is necessary to use the Transformation Method Theorem and the independence between the times and the angles: 

$$\displaystyle f_{S_{N_{j}-1}}(s_1)=\int f_{V_{N_{j}}}(\frac{s_1}{s_2})f_{\tilde W_{N_{j}-1}}(s_2) I_{\{-s_2\leq s_1\leq w_2\}}(s_1) I_{\{s_2\leq-c_j\}}(s_2) \cdot ds_2$$

\subsection*{Summary Statistics}
\begin{figure}
  \includegraphics[width=0.8\textwidth]{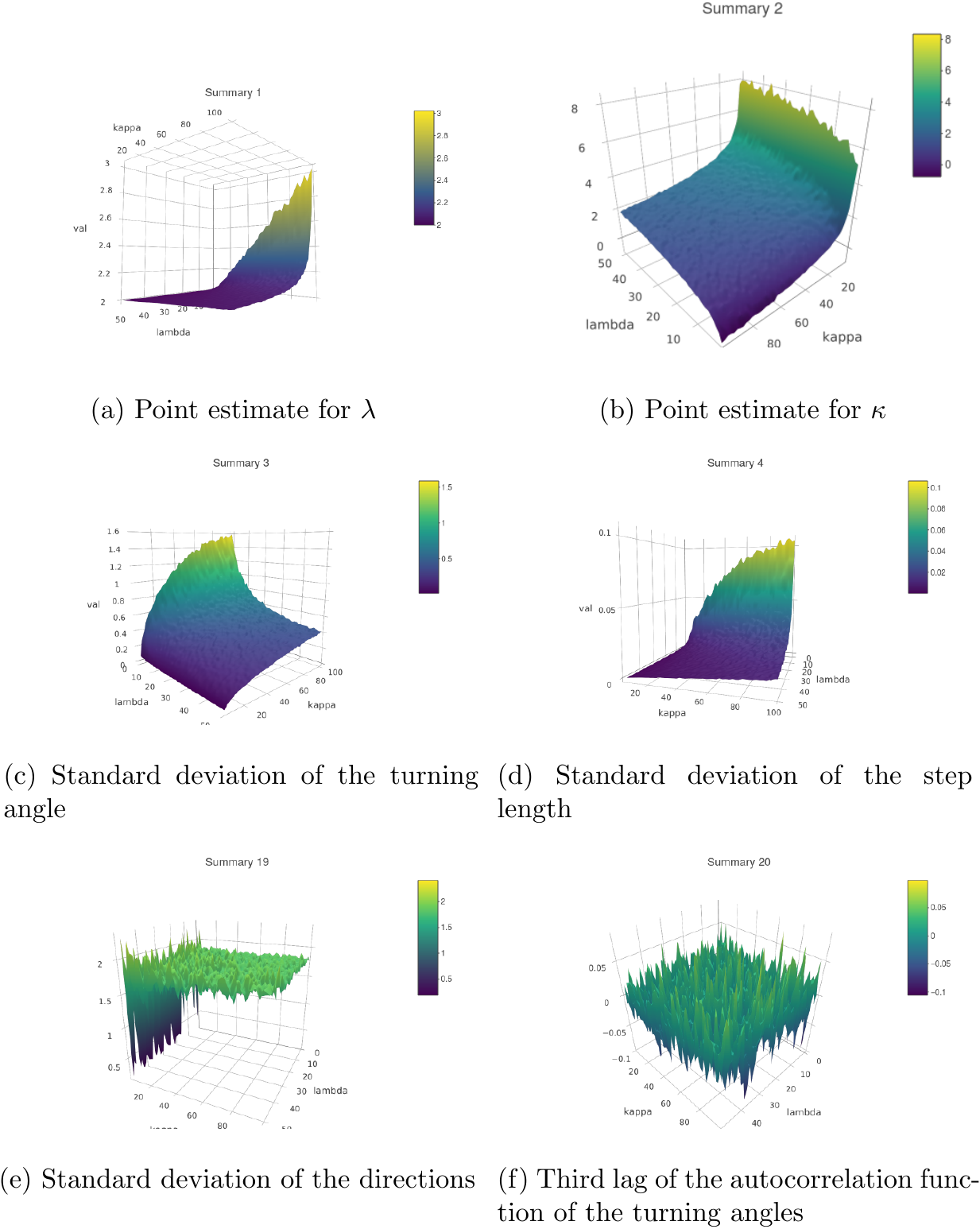}\label{summ1}
\caption{Plots of the summary statistics vs the simulated parameters.}
\label{PLotsummaries1}
\end{figure}

\begin{figure}[H]
\centering

  \includegraphics[width=0.8\textwidth]{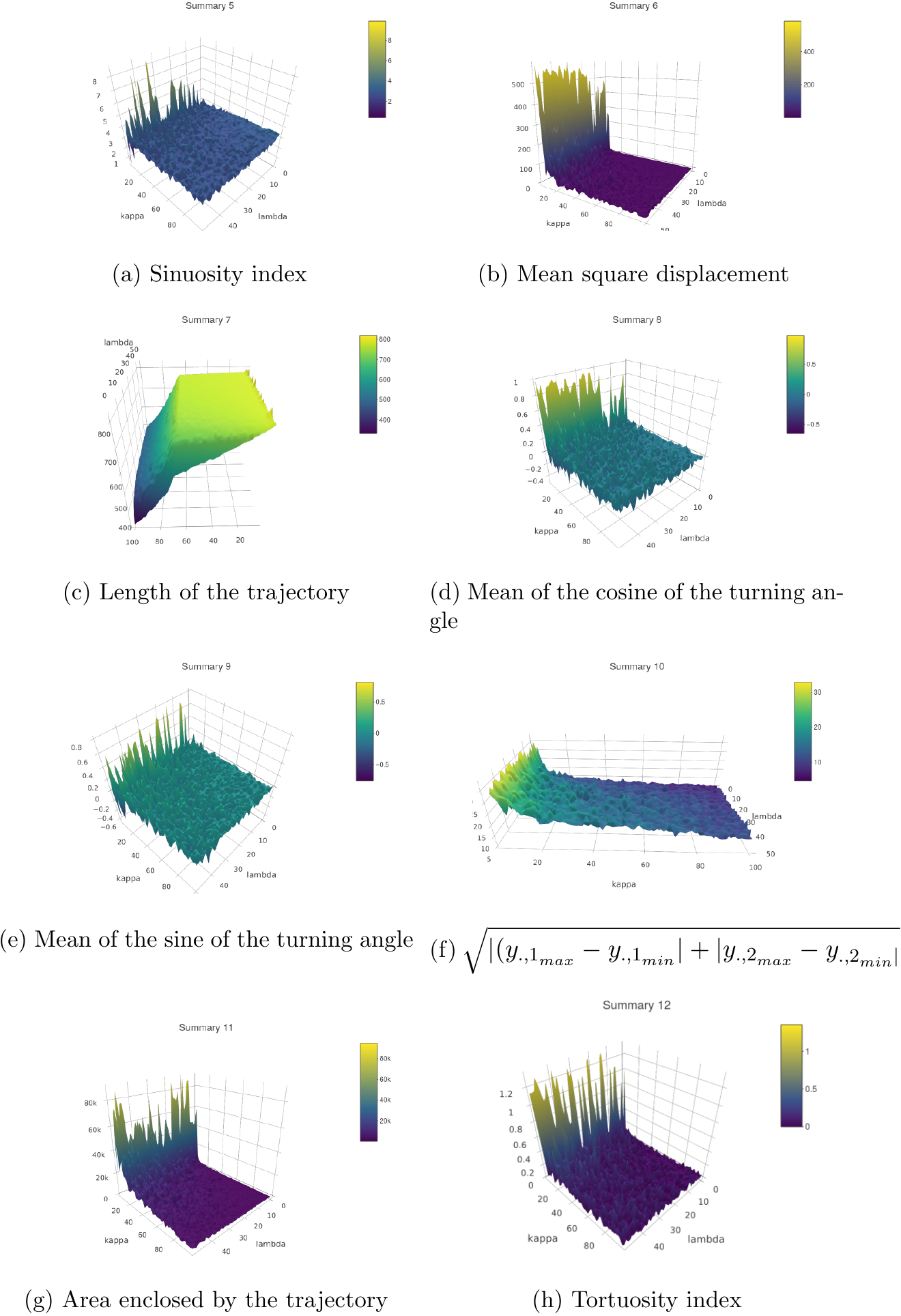}\label{summ5}

\caption{Plots of the summary statistics vs the simulated parameters.}

\label{PLotsummaries2}
\end{figure}

\begin{figure}[H]
\centering

  \includegraphics[width=0.8\textwidth]{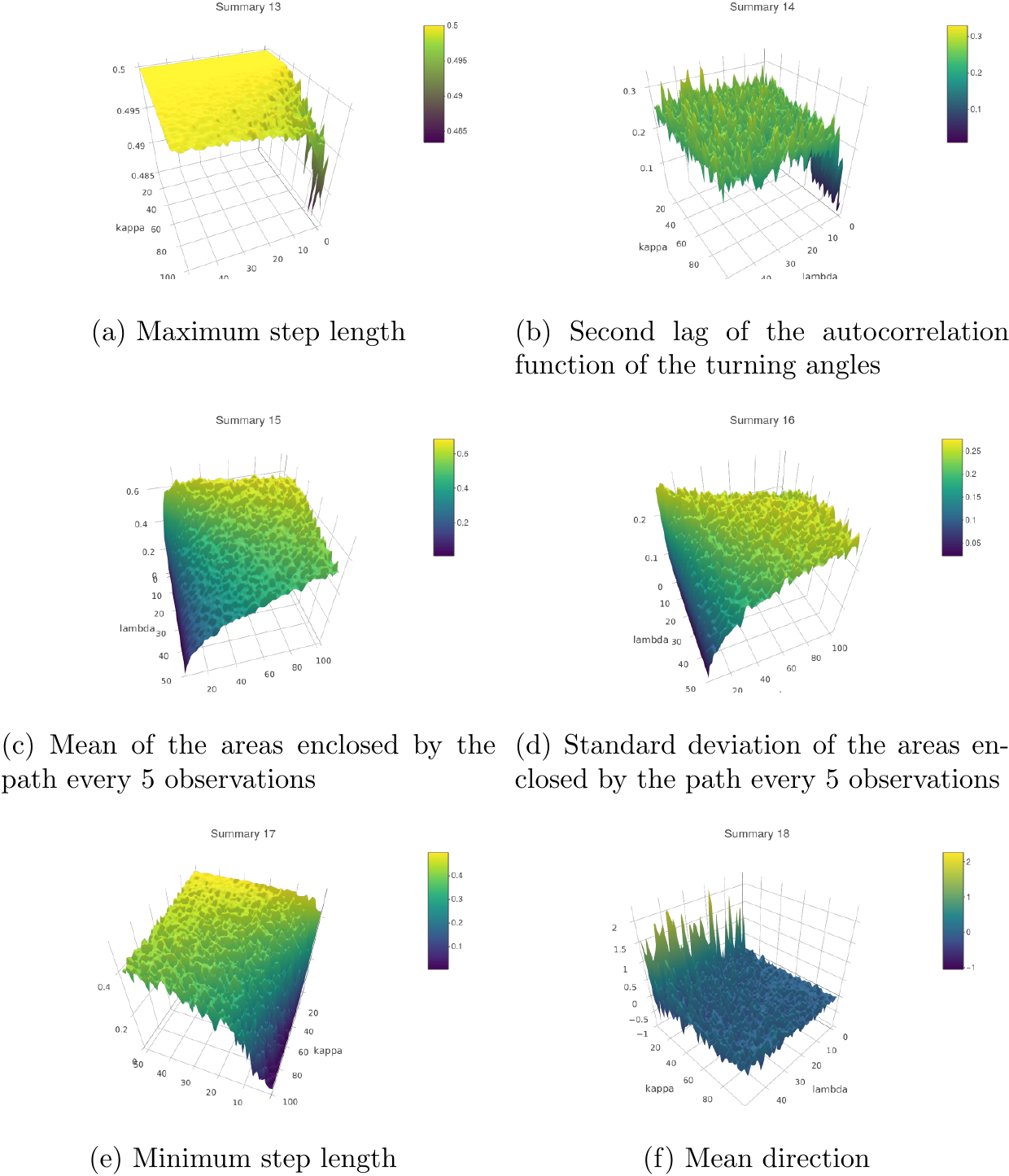}
\caption{Plots of the summary statistics vs the simulated parameters.}
\label{PLotsummaries3}
\end{figure}

We provide the plots of the summary statistic analyzed versus the parameters (Figures \ref{PLotsummaries1},
\ref{PLotsummaries2}, and \ref{PLotsummaries3}).
 We choose four that attempt to describe the trajectories in an integral way and characterize them according to parameter values.
 The selected were \ref{PLotsummaries1}(a), \ref{PLotsummaries1}(b),\ref{PLotsummaries1}(c) and \ref{PLotsummaries1}(d)

\subsection*{Empirical Coverage}

We present the results for the empirical coverage of the $95\%$ high posterior density(HPD) intervals for the two parameters. This value is the proportion of simulations for which the true parameter value falls within the $95\%$ HPD interval. 
If the posterior distributions were 
correctly estimated, this proportion should have been near $0.95$. 
We compute this index for both parameters ($\kappa$ and $\lambda$) and for the three ABC algorithms: Simple Rejection, Corrected via Linear Regression and Corrected via Neural Network. We did that for threshold ($\epsilon$) values of: $0.001$, $0.005$, $0.01$ and $0.1$. 

\begin{figure}[H]
\centering
\includegraphics[scale=0.5]{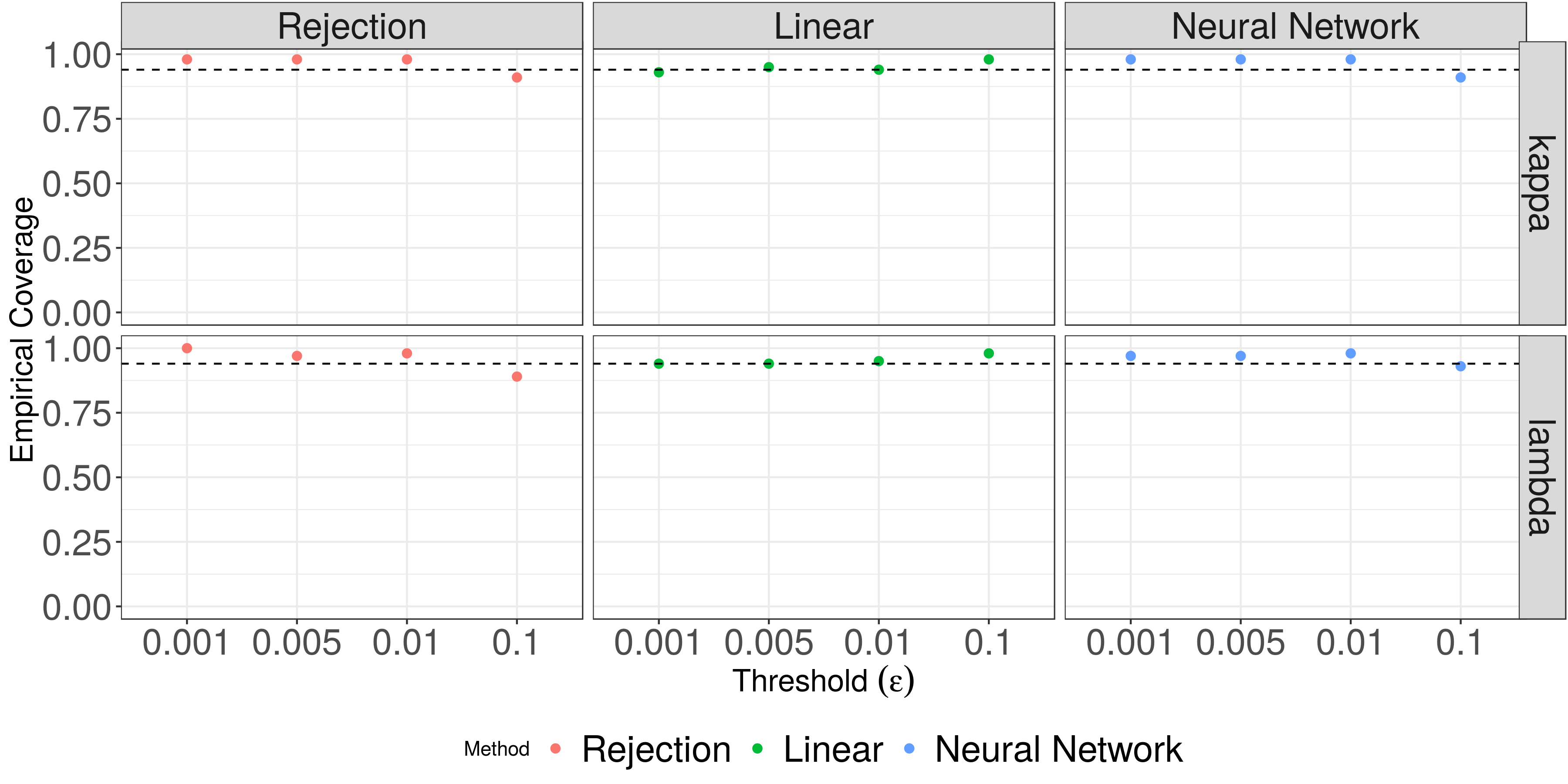}
\caption{Empirical Coverage obtained for $\kappa$ and $\lambda$ for different threshold values. By row the results for every parameter and by column the results for the three ABC algorithms. The dash line indicates the value $0.95$ } \label{ecCV}
\end{figure}

\end{document}